\documentclass[12pt, a4paper]{article}

\usepackage[
  margin=0.75in,
  headsep=10pt, 
]{geometry}

\usepackage{graphicx}
\usepackage{soul}
\usepackage{graphicx}
\usepackage{float}
\usepackage{epstopdf, epsfig}
\usepackage{amsmath}
\usepackage[table]{xcolor}
\usepackage{subcaption}
\usepackage{soul}
\usepackage{makecell}
\usepackage{multirow}
\usepackage{breqn}
\usepackage{amssymb} 
\usepackage{booktabs}
\usepackage{dcolumn}
\usepackage{bm}
\usepackage[utf8]{inputenc}
\usepackage[T1]{fontenc}
\usepackage{mathptmx}
\usepackage{etoolbox}
\usepackage{hyperref}
\usepackage{subcaption}
\usepackage{ragged2e}
\usepackage[sort&compress]{natbib} 

\newcommand{\edt}[1]{{\color{black}#1}} 
\newcommand{\fm}[1]{{\color{black}#1}} 

\newcommand{\soptitle}{Design Implications of Chord Length and Number of Blades on Self-Starting Process in Vertical-Axis Wind Turbines}
\usepackage{xcolor} 

\begin{document}


\begin{center}
\Large \bf{\soptitle}
\vspace{0.1in}
\end{center}

\begin{center}
{Faisal Muhammad$^1$, and Muhammad Saif Ullah Khalid$^{1\ast}$}\\
\vspace{0.1in}
\end{center}
\begin{center}
$^1$Nature-Inspired Engineering Research Lab (NIERL), Department of Mechanical and Mechatronics Engineering, Lakehead University, Thunder Bay, ON P7B 5E1, Canada\\
\vspace{0.05in}
$^\ast${Corresponding Author, Email: mkhalid7@lakeheadu.ca}
\end{center} 

\begin{abstract}

Self-starting remains a key limitation of lift-driven vertical-axis wind turbines and is strongly influenced by geometric design choices that also govern steady-state performance. This work quantifies the roles of chord length and blade number on startup dynamics and the attained steady tip-speed ratio using two-dimensional URANS simulations of freely rotating Darrieus-type rotors. Two configuration families are examined, an equal-chord set in which three and five bladed turbines share the same chord length, and an equal-solidity set in which the chord length is reduced for the five blade turbines to match solidity with the three blade counterparts. Results are analyzed using the time evolution of tip-speed ratio, reduced-frequency measures to identify sustained unsteady intervals, vorticity-field diagnostics of dynamic stall vortex formation and detachment, and a torque decomposition into pressure and viscous moments. The results show that increasing number of blades can enhance early stage acceleration but generally lowers the steady tip-speed ratio by intensifying blade-vortex interaction in the downstream half cycle. Increasing chord length promotes self-starting by strengthening unsteady loading during the transition out of the low-speed regime, but also increases viscous losses and wake interaction, leading to lower the steady tip-speed ratio for self-starting high chord configurations. The role of viscous moments is also analyzed to quantify their contribution to self-starting behavior and to assess their influence on limiting the attainable operating state after self-starting. These findings provide design-relevant guidance on the startup--performance trade-off associated with $c$ and $N$ in freely accelerating VAWTs.


\end{abstract}

\section{Introduction}
\label{sec:Intro}

Wind turbines can broadly be categorized into \edt{horizontal-axis wind turbines} (HAWTs) and \edt{vertical-axis wind turbines} (VAWTs) based on \edt{the} rotational axis \edt{of their blades} \citep{paraschivoiu2002wind}, each with distinct design characteristics and operational advantages. HAWTs, the more common type used in large-scale wind farms, offer high efficiency in open, unidirectional wind conditions due to their ability to align with the prevailing direction of wind. However, they require yaw mechanisms to track the wind, tall towers for mounting, and large clearances for blade rotation \citep{johari2018comparison, mendoza2019performance}), increasing \edt{complexity and cost associated with installations}. Moreover\edt{,} HAWTs also pose a risk to avians and have vast geo-political and electro-magnetic interference impact \citep{kumar2018critical}. \edt{Contrarily}, VAWTs rotate around a vertical axis and can function through the wind from all directions without reorientation, making them especially suitable for turbulent urban wind environments, lower wind speeds, and decentralized applications. Their ground-level generators simplify maintenance, and their compact design enables integration into rooftops, highways, and other constrained locations. Despite these advantages, VAWTs generally exhibit lower aerodynamic efficiency and have faced challenges with self-starting and dynamic stability.

\begin{figure}[H]
  \centering
  \includegraphics[width=0.75\linewidth]{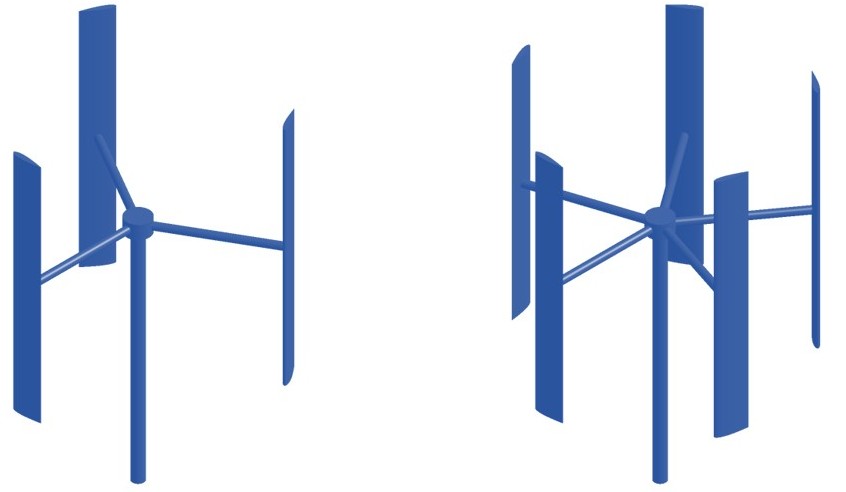}
  \caption{\edt{Representative models} of \edt{three-} and \edt{five-bladed rotors}}
  \label{fig:sch}
\end{figure}

Among VAWTs, two main types dominate, \edt{including} Savonius and Darrieus turbines. The Savonius turbine is a drag-based design that offers excellent self-starting capabilities, simplicity, and robustness at lower angular velocities, but suffers from low efficiency due to significant drag forces. It is commonly used in low-power or remote applications\edt{,} such as ventilation or water pumping \citep{fanel2021review}. On the other hand, the Darrieus turbine is a lift-based design that leverages aerodynamic forces to achieve much higher efficiency at moderate to high \edt{tip-speeds ratios} ($\lambda$). However, traditional Darrieus turbines are not inherently self-starting and may experience torque ripples and dynamic instabilities at certain operating ranges. \edt{Two representative models of VAWTs with three and five blades are shown in Fig.~\ref{fig:sch}}. 

\begin{figure}[H]
  \centering
  \includegraphics[width=0.8\linewidth]{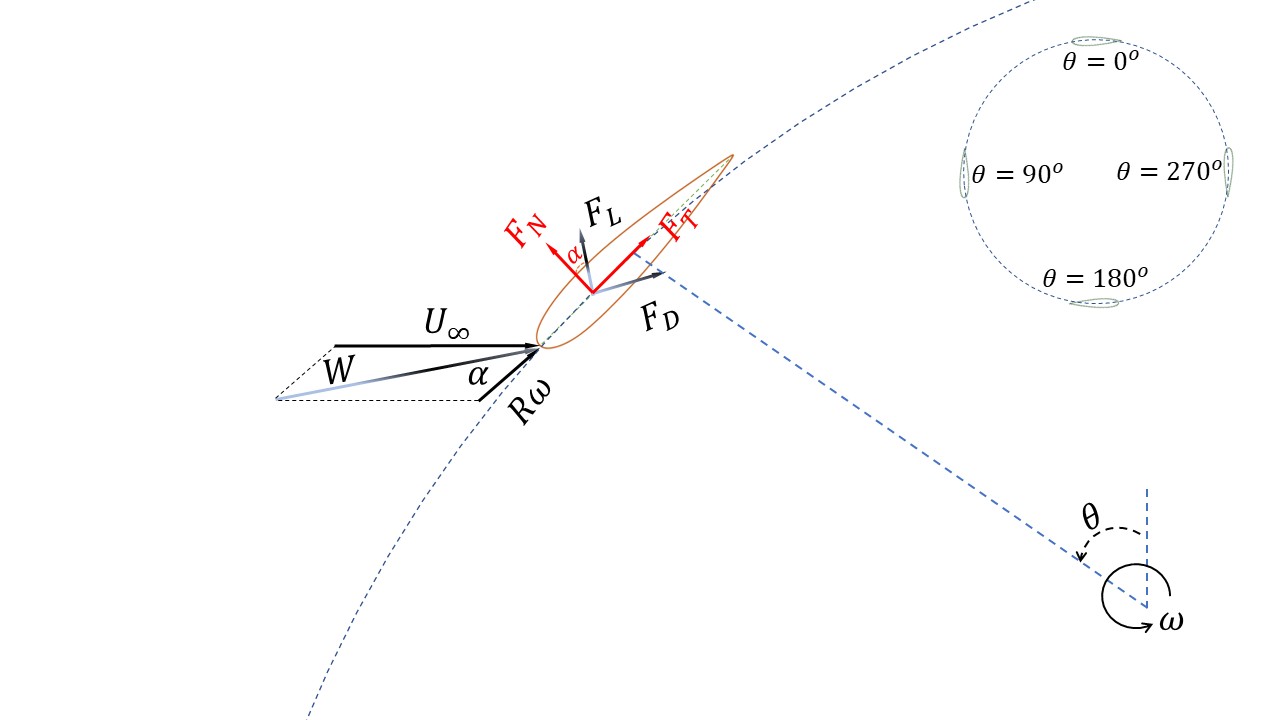}
  \caption{Characteristic \edt{v}elocities and \edt{f}orces acting \edt{on a blade} in a VAWT}
  \label{fig:charac}
\end{figure}

\edt{During the whole rotational} cycle, the blades of a VAWT experience strongly unsteady flow conditions. To illustrate the self-starting mechanism, consider a single non-pitching blade rotating counter-clockwise with angular velocity $\omega$, as shown in Fig.~\ref{fig:charac}. The instantaneous position of the blade is given by the azimuthal angle ($\theta$), measured from the \edt{vertically} upward direction. The vectorial addition of free-stream velocity ($U_\infty$) and the tangential velocity ($R\omega$) form the effective velocity vector ($U_{eff}$) experienced by the blade. \edt{Here, $R$ and $\omega$ denote the radius and angular speed of the turbine, respectively.} The magnitude \edt{of $U_{eff}$} and \edt{its} effective instantaneous angle ($\alpha_{eff}$) with the chord can be expressed as \fm{(\cite{le2024optimal, zhu2019numerical,posa2021secondary})}\edt{:} 

\begin{equation}
U_{eff} = U_\infty \sqrt{\lambda^2 + 2\lambda \cos\theta + 1},
\label{eq:Wmag}
\end{equation}

\begin{equation}
\alpha_{eff} = \tan^{-1}\!\left(\frac{\sin\theta}{\cos\theta + \lambda}\right).
\label{eq:alpha}
\end{equation}

\noindent where $\theta$ is measured positive in the anti-clockwise direction starting from the top position ($\theta=0$), as illustrated in Fig.~\ref{fig:charac}, and $\lambda$ is the instantaneous tip-speed ratio, defined as\edt{:}

\begin{equation}
\lambda = \frac{R\,\omega}{U_\infty},
\label{eq:tsr}
\end{equation}

The geometric loading of the rotor is described by solidity \edt{(}$\sigma$\edt{)}, which is \edt{defined} by the following expression\edt{:}

\begin{equation}
\sigma = \frac{N\,c}{R},
\label{eq:sol}
\end{equation}

\edt{\noindent where $N$ shows the number of blades in a rotor, and $c$ is the chord-length of a blade.}

At lower $\lambda$, $\omega$ (\edt{and the resultant tangential velocity of the blade}) is small relative to $U_\infty$, so the relative flow is dominated by the free stream. As illustrated in Fig.~\ref{fig:charac}, \edt{it} yields a comparatively large $\alpha_{\mathrm{eff}}$ \edt{and its large variations with time during a rotational cycle}. As the rotor accelerates, \edt{the tangential velocity of a blade} increases and contributes more strongly to the relative velocity, causing the relative velocity vector to \edt{orient more} toward the \edt{blade's kinematics} and progressively reducing $\alpha_{\mathrm{eff}}$. Consequently, the range of $\alpha_{\mathrm{eff}}$ \edt{gets narrowed} as $\lambda$ increases, leading to a more stable aerodynamic response during the subsequent acceleration and steady operating stages. In the literature \fm{\citep{zhu2022numerical,kong2024improvement,zare2024self}}, the term self-starting \edt{was} used with varying interpretations. In the present work, it is taken to refer to the ability of a turbine to accelerate autonomously from rest and reach a steady operating state in which the \edt{blade's} tip speed exceeds the wind speed. A VAWT is therefore considered to have self-started once it reaches $\lambda > 1$, at which point sustained rotation is established \citep{khalid2022self,takao2009straight}.

One important phenomenon associated with VAWTs that strongly influences their self-starting characteristics is dynamic stall. First reported in the context of helicopter aerodynamics \citep{ham1968dynamic}, dynamic stall \edt{received} substantial attention in the VAWT\edt{-related} literature\edt{,} because \edt{their} start-up \edt{process} frequently occurs in a low-$\lambda$ regime\edt{,} where the blades experience large, rapidly varying $\alpha_{\mathrm{eff}}$ over the azimuthal cycle. The formation and shedding of \edt{large coherent flow structures cause} intense unsteady aerodynamic loading \edt{on the blades. It} can be detrimental to overall performance \edt{of the turbine} and can also \edt{adversely} affect \edt{its} structural integrity by \edt{causing} fatigue and triggering aeroelastic instabilities\edt{,} \textit{e.g.}, flutter, potentially leading to failure of \edt{the turbine's} components \citep{de2021controlling,dunne2015dynamic,ouro2018effect}. 

\fm{In the context of VAWTs, the initial operating stage, commonly termed the startup phase, occurs in low speed regimes (low-$\lambda$), where turbine geometry strongly governs the aerodynamic response. A key geometric parameter is $\sigma$, typically interpreted as the ratio of total blade planform area to the rotor's swept area \citep{miller2021solidity}. Prior work indicates that the self starting capability of Darrieus type VAWTs generally improves with increasing $\sigma$ \citep{mohamed2013impacts}. However, low-$\lambda$ operation also places the turbine at an elevated risk of dynamic stall, motivating design and control strategies to mitigate stall induced losses and unsteady loading. \edt{For instance}, variable pitching of blades was recommended in low-$\lambda$ conditions by \fm{Sagharichi et al.} \cite{sagharichi2018effect}, \edt{reporting} a variable pitch turbine with $\sigma \approx 0.5$ \edt{operated} in regimes prone to dynamic stall. \edt{Besides single-rotor VAWTs}, \edt{experimental} results for dual VAWT configurations \edt{showed} that \edt{changes in the produced power could} be weak at low $\lambda$, while lower $\sigma$ \edt{increased} sensitivity \edt{spacing between the rotors and skewness in the flow} \citep{li2021experimental}. \edt{It implies} that urban or array \edt{configurations} should account for solidity, spacing, and skewed inflow effects. Complementing these findings, experimental data \citep{mclaren2011numerical} for high-$\sigma$ turbines ($\sigma \approx 0.42$ to $0.45$) at low $\lambda$ indicate that the startup regime is dominated by deep dynamic stall and strong unsteady loading, \edt{where} accurate predictions may require \edt{two-dimensional modeling through unsteady Reynolds-Averaged Navier-Stoke Equations (URANs)} augmented with simple correction factors \edt{to work with three-dimensional systems}. Moreover, Tong et al. \cite{tong2023effects} \edt{reported that the} chord\edt{-}to\edt{-}circumference ratio was an additional key design parameter for small-bladed VAWTs rated below $10~\mbox{kW}$. \edt{It reveals} inherent trade offs between starting capability and power performance as \edt{the} rotor\edt{'s} diameter and chord length were varied. Collectively, these \edt{investigations} indicate that geometric parameters, including \edt{the blade's} chord length, rotor\edt{'s} diameter, and related quantities embedded within solidity\edt{-}driven design, \edt{exhibit} a strong influence on startup behavior and unsteady aerodynamics of VAWTs. \edt{However, significant gaps exist related to suitable geometric parameters for self-starting of VAWTs in the current literature. Our present work addresses this important aspect of designs of VAWTs, which systematically examines the influence of chord-length and number of blades, independent of solidity, to characterize their roles on the startup dynamics, dynamic stall, and the eventual stead-state $\lambda$.} \edt{Therefore, the novelty of this work revolves around addressing the following important research questions that are unanswered in literature, according to the authors' knowledge: (i} how does the turbines with different number of blades but with the same chord of each blade or same solidity of their rotors self start? (ii) what are the aerodynamic indicators for investigating the reasons behind their ability or inability to self start under different operating conditions? (iii) what is the role of flow unsteadiness in governing the onset of self-starting, dead-band behavior, and transition to sustained rotation in freely accelerating VAWTs? (iv) how do viscous effects, through their contribution to the aerodynamic moment, influence the self-starting process and limit or support the attainable operating state after startup?} 

\fm{To address these research questions, the present study examines} how two geometric features, $c$ and $N$, influence the start-up response of VAWTs. The motivation for this research work arises from the limited literature that explicitly couples variations in $c$ and $N$. \edt{These} are often discussed only through the solidity, even though solidity is primarily a geometric descriptor rather than a similarity parameter for the start-up dynamics \edt{\citep{wei2021near}. The explanation of our work \edt{begins} with describing the computational} setup and geometric configurations in \edt{s}ection~\ref{sec:NumVal}. The results are presented in \edt{s}ection~\ref{sec:resrults} \fm{by systematically addressing the research questions posed earlier}, beginning with a summary of the overall start-up outcomes and then relating the observed trends to the underlying aerodynamic mechanisms and their implications for both start-up and steady operating performance. Finally, the main conclusions are provided in \edt{S}ection~\ref{sec:conclusions}.

\section{\edt{Simulation Framework}}
\label{sec:NumVal}
We present the methodology adopted for the numerical simulations \edt{for flows around a range of configurations of VAWTs} in this section, starting with \edt{the geometry and domain of the simulations (section~\ref{subsec:Geom}), followed by \edt{explanation of} the numerical solver (section~\ref{subsec:CM})}. \edt{Then, we present the results of the} mesh\edt{-convergence} and time-step independence studies in section~\ref{subsec:Space} \edt{with details of the validation cases in section}~\ref{subsec:val}. 

\subsection{\edt{Geometry of VAWTs and the Computational Domain}}
\label{subsec:Geom}

The \edt{primary geometric configuration} of VAWTs considered in this study \edt{is inspired from the one used by} \fm{Rainbird} \cite{rainbird2007aerodynamic} in their experimental investigations. \edt{The other geometries are based on the same VAWT, as it was also extensively employed in other studies reported in recent} literature \citep{khalid2022self, asr2016study, fatahian2024optimization,bianchini2016experimental, celik2022design}. To investigate the influence of $\sigma$ on the self-starting behavior \edt{of VAWTs}, only $c$ and $N$ are varied, while $R$ is kept constant for all configurations. Two configuration sets are considered, the equal\edt{-}chord ($\mbox{EC}$) set and \edt{another} equal\edt{-}solidity ($\mbox{ES}$) set, as summarized in Table~\ref{tab:geom}.

In the $\mbox{EC}$ set, \edt{$3$-bladed} and \edt{$5$-bladed} configurations share identical chord lengths, such that \edt{the} increasing $N$ leads to higher $\sigma$. In the $\mbox{ES}$ set, the chord \edt{of each blade} of the \edt{$5$-bladed} configurations is reduced\edt{,} such that \edt{their} $\sigma$ matches that of the corresponding \edt{$3$-bladed system}, isolating the effect of $N$ at \edt{a} fixed $\sigma$. This coordinated design enables the independent assessment of the roles of $c$, $N$, and $\sigma$ on the self-starting \edt{behavior of the turbines}.

\begin{table}[h!]
\centering
\caption{Geometric parameters of \edt{$3$-bladed} and \edt{$5$-bladed} VAWTs\edt{, where the subscripts for $c$ and $\sigma$ represent the corresponding number of blades in a turbines}}
\begin{tabular}{lcc}
\hline
\textbf{Parameter} & \textbf{3 blades} & \textbf{5 blades} \\
\hline
Set 1 $c$ (mm) $c_3$=$c_5$  & 43, 63, 83, 103, 123 & 43, 63, 83, 103, 123 \\
Set 2 $c$ (mm) $\sigma_3$=$\sigma_5$ & 43, 63, 83, 103, 123 & 25.8, 37.8, 49.6, 61.6, 73.6 \\
$R$ (m)        & 0.375                & 0.375 \\
$H$ (m)        & 1                    & 1 \\
Blade profile   & NACA0018             & NACA0018 \\
Moment of inertia (kg\,m$^2$) & 0.018 & 0.018 \\
\hline
\label{tab:geom}
\end{tabular}
\end{table}

To mitigate influence \edt{of boundaries of the computational domain} on performance of the turbine and alleviate effects of blockage on the \edt{numerical} results, we follow the practices recommended by \fm{Rezaeiha et al.} \cite{rezaeiha2017cfd}. The \edt{center of the} turbine is placed \edt{at a distance of} $10D$ from the inlet, top\edt{,} and bottom boundaries, whereas the outlet is placed at a distance of $15D$. 

\subsection{Computational Method}
\label{subsec:CM}
This study employs OpenFOAM, an open-source finite-volume solver widely used for wind-turbine investigations (\cite{zhang2019rans, stovall2010wind, chu2017biomimetic,kim2025high,ricci2025review}). The incompressible unsteady Reynolds-averaged Navier–Stokes (URANS) equations (\fm{Eq.~\ref{eq:mom_urans}}) for wind around the turbine are solved using a pressure-based, segregated algorithm. In this work\edt{,} we use pimpleFoam, which merges PISO (Pressure Implicit Splitting of Operators) and SIMPLE (Semi-Implicit Method for Pressure-Linked Equations) into the PIMPLE procedure (outer SIMPLE-style loops with inner PISO-style correctors, \cite{nguyen2025openfoam}). PimpleFoam is expressly formulated as a large \edt{timestep-based} transient solver for incompressible flows \cite{penttinen2011pimplefoam}, \edt{relying on} multiple outer corrections per \edt{timestep} to maintain stability\edt{, accuracy, and efficiency of simulations for scenarios, such as flows over rotating VAWTs}. \edt{Here, the} spatial \edt{convective and diffusive terms are discretized using the} second\edt{-}order least-squares/linear gradients techniques, and the transient term \edt{is approximated via} a second-order implicit scheme. \edt{The angular motion of a VAWT} is handled via \edt{the} sliding-mesh {technique}, which permits \edt{the rotor to rotate physically} without \edt{requiring any} re-meshing \edt{in the computational domain}. \edt{Due to high-Reynolds number flows involved here, we model turbulence using} the $k$–$\omega$ SST model augmented by the $\gamma$–$\mathrm{Re}_\theta$ transition \cite{zarmehri2012gamma} to capture laminar–turbulent transition pertinent to \edt{situations encountered by VAWTs}. 

\fm{The incompressible URANS continuity and momentum equations solved in this work are
\begin{align}
\nabla\cdot\overline{\mathbf{U}} &= 0, \label{eq:cont_urans}\\
\rho\left(\frac{\partial \overline{\mathbf{U}}}{\partial t}+(\overline{\mathbf{U}}\cdot\nabla)\overline{\mathbf{U}}\right)
&= -\nabla \overline{p}
+\nabla\cdot\left[(\mu+\mu_t)\left(\nabla\overline{\mathbf{U}}+(\nabla\overline{\mathbf{U}})^{T}\right)
-\frac{2}{3}\rho k\,\mathbf{I}\right]. \label{eq:mom_urans}
\end{align}
Here, $\overline{\mathbf{U}}$ and $\overline{p}$ are the Reynolds-averaged velocity and pressure, $\rho$ is the fluid density,
$\mu$ is the dynamic viscosity, $\mu_t$ is the eddy viscosity, $k$ is the turbulent kinetic energy (appearing through the isotropic
part of the Reynolds stresses), and $\mathbf{I}$ is the identity tensor.

Turbulence closure is provided by the $k$--$\omega$ SST model, for which the transport equations are
\begin{align}
\frac{\partial (\rho k)}{\partial t}+\nabla\cdot(\rho k\,\overline{\mathbf{U}})
&= P_k-\beta^{*}\rho k\omega+\nabla\cdot\left[(\mu+\sigma_k\mu_t)\nabla k\right], \label{eq:k_sst}\\
\frac{\partial (\rho \omega)}{\partial t}+\nabla\cdot(\rho \omega\,\overline{\mathbf{U}})
&= \alpha\,\rho\frac{P_k}{k}-\beta\,\rho\omega^2
+\nabla\cdot\left[(\mu+\sigma_\omega\mu_t)\nabla \omega\right]
+2(1-F_1)\rho\sigma_{\omega2}\frac{1}{\omega}\,\nabla k\cdot\nabla\omega. \label{eq:omega_sst}
\end{align}
In Eqs.~\eqref{eq:k_sst}--\eqref{eq:omega_sst}, $k$ is the turbulent kinetic energy, $\omega$ is the specific dissipation rate,
$P_k$ is the production of $k$, and $F_1$ is the SST blending function.
}

\subsection{\edt{Grid-Convergence and Timestep-Independence Tests}}
\label{subsec:Space}
\begin{figure}[H]
  \centering
  \includegraphics[width=1\linewidth]{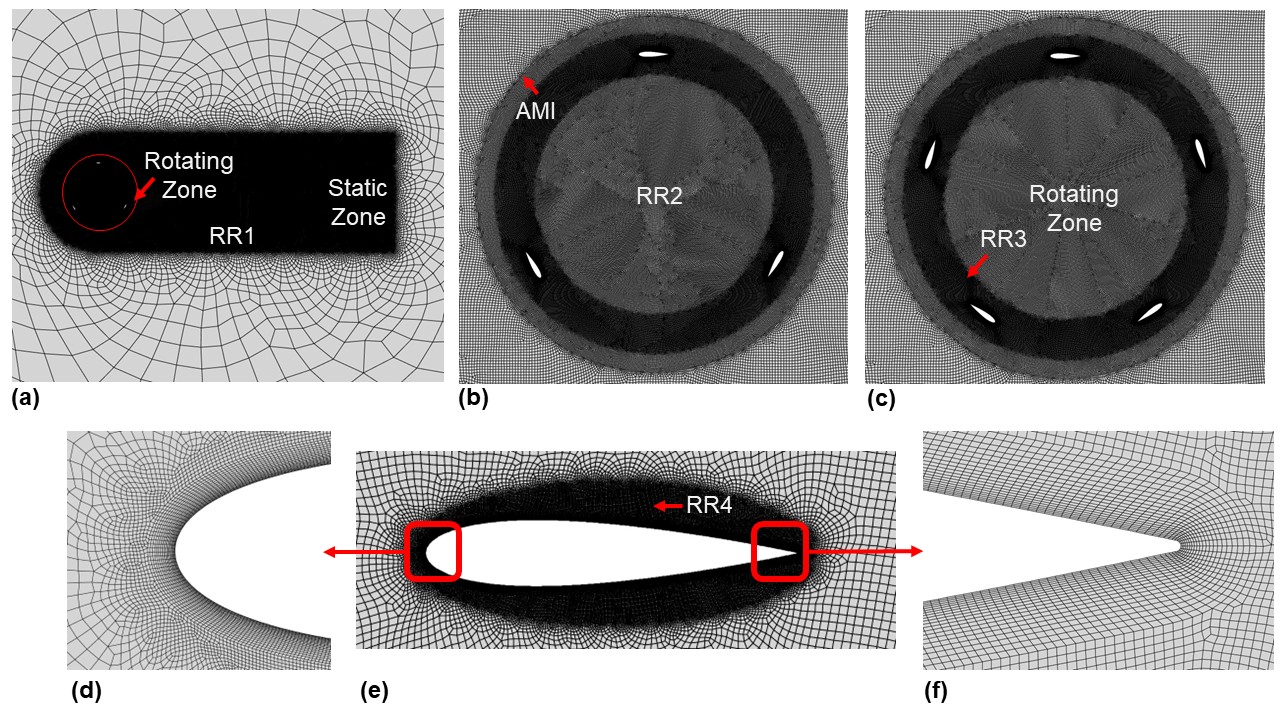}
  \caption{Nested refinement regions inside the static and rotating zones (RR1--RR4)}
  \label{fig:rotor-refine}
\end{figure}

\edt{The computational domain around a turbine is discretized through} unstructured, quad-dominant mesh, \edt{where} a hierarchical refinement strategy is applied to obtain grid-independent solutions. The domain is partitioned into a static zone and a rotating zone with multiple refinement levels. In the static zone, the wake is refined as \edt{the} refinement region 1 ($\mbox{RR1}$), extending to \edt{a downstream distance of} $5D$ \edt{from the center of the turbine} and \edt{to} $2D$ \edt{in the vertical direction,} as shown in Fig.~\ref{fig:rotor-refine}(a). Within the rotating zone, \edt{we establish} three refinements, \edt{including} $\mbox{RR2}$ \edt{as the refinement in the core volume} (see Fig.~\ref{fig:rotor-refine}b), $\mbox{RR3}$ \edt{as} an annular band along the \edt{trajectory of a blade} (see Fig.~\ref{fig:rotor-refine}c), and $\mbox{RR4}$ \edt{as} an elliptical sheath enveloping each blade, with \edt{the enlarged view of the} leading and trailing \edt{edges} shown in Fig.~\ref{fig:rotor-refine}d and Fig.~\ref{fig:rotor-refine}f, \edt{respectively}. \edt{Its} overall extent is \edt{presnted in Fig.~\ref{fig:rotor-refine}e}. To resolve the viscous sub-layers of the boundaries of the blades, inflated quadrilateral cells are introduced with a total of $20$ layers \edt{of grid cells} from the wall\edt{,} as suggested by \fm{Ramirez and Saravia} \cite{ramirez2021assessment}. \edt{The} first\edt{-cell} height \edt{is chosen based on} \edt{the value of} $y^+$ \edt{as $5$} \edt{stretching the grid with a ratio of $1.08$, following the recommendation} by \fm{Li et al.} \cite{li20132}. 


\begin{figure}[h!]
  \centering
  \includegraphics[width=0.8\linewidth]{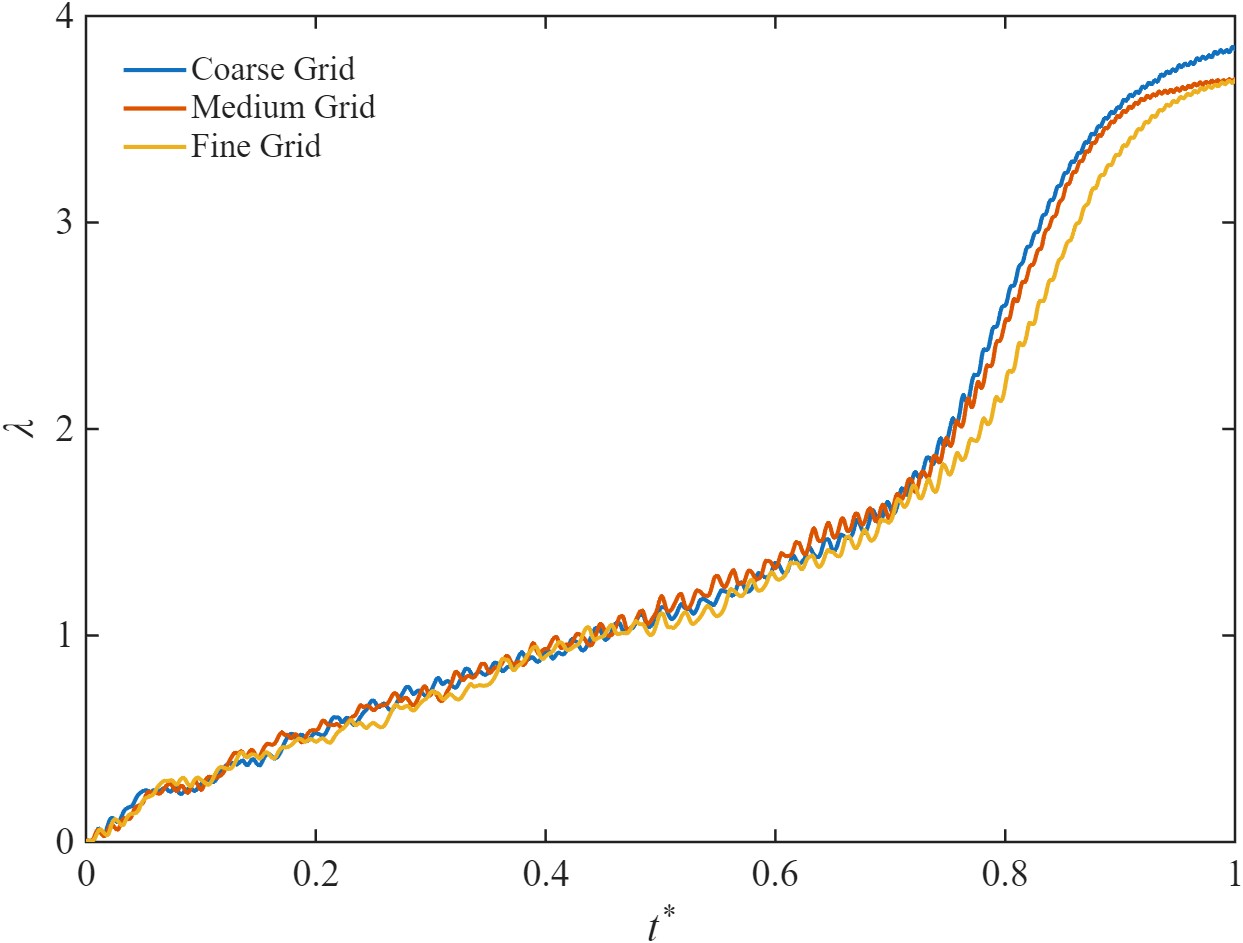}
  \caption{\edt{Comparison of tip-speed ratios of a VAWT using three different grids}}
  \label{fig:GridInd}
\end{figure}

 We carry out the \edt{grid-convergence} and \edt{timestep-independence tests} on a \edt{3-bladed} turbine with an inlet velocity of $8~\mbox{m/s}$, \edt{where} the \edt{rotor's} moment of inertia is kept at 0.018~\(\mathrm{kg\,m^2}\), \fm{following the benchmark configuration reported by \fm{Rainbird} \cite{rainbird2007aerodynamic}}. Three meshes \edt{with} coarse, medium, and fine \edt{resolutions} are generated for the grid-independence study, and the cell sizes in $\mbox{RR1}$ to $\mbox{RR4}$ are varied systematically\edt{,} as shown in Table~\ref{tab:refinement-sizes}. \edt{Figure~\ref{fig:GridInd} presents the tip-speed ratios versus time, obtained using the three mesh configurations to compare the self-staring behavior of the turbine. It is evident that} the medium and fine grids exhibit nearly identical behavior and converge to the same steady state $\lambda$. \edt{, following the practices reported by \fm{Asr et al.} \cite{asr2016study} and \fm{Khalid et al.} \cite{khalid2022self},} \(t^*\) denotes the total time normalized by the time required for the turbine to reach its steady-state $\lambda$ \edt{here}. 

\begin{table}[h!]
\centering
\caption{Refinement sizes in each region and total cell counts.}
\label{tab:refinement-sizes}
\begin{tabular}{lccccr}
\hline
Refinement Level & RR1 (m) & RR2 (m) & RR3 (m) & RR4 (m) & No. of cells \\
                 & $\times 10^{-3}$ & $\times 10^{-3}$ & $\times 10^{-3}$ & $\times 10^{-4}$ & \\
\hline
Coarse & 9   & 7   & 5   & 3.5 & 163{,}000 \\
Medium & 7.5 & 3.5 & 2   & 2.5 & 312{,}000 \\
Fine   & 6   & 2   & 1.5 & 2   & 540{,}000 \\
\hline
\end{tabular}
\end{table}

\begin{figure}[H]
  \centering
  \includegraphics[width=1\linewidth]{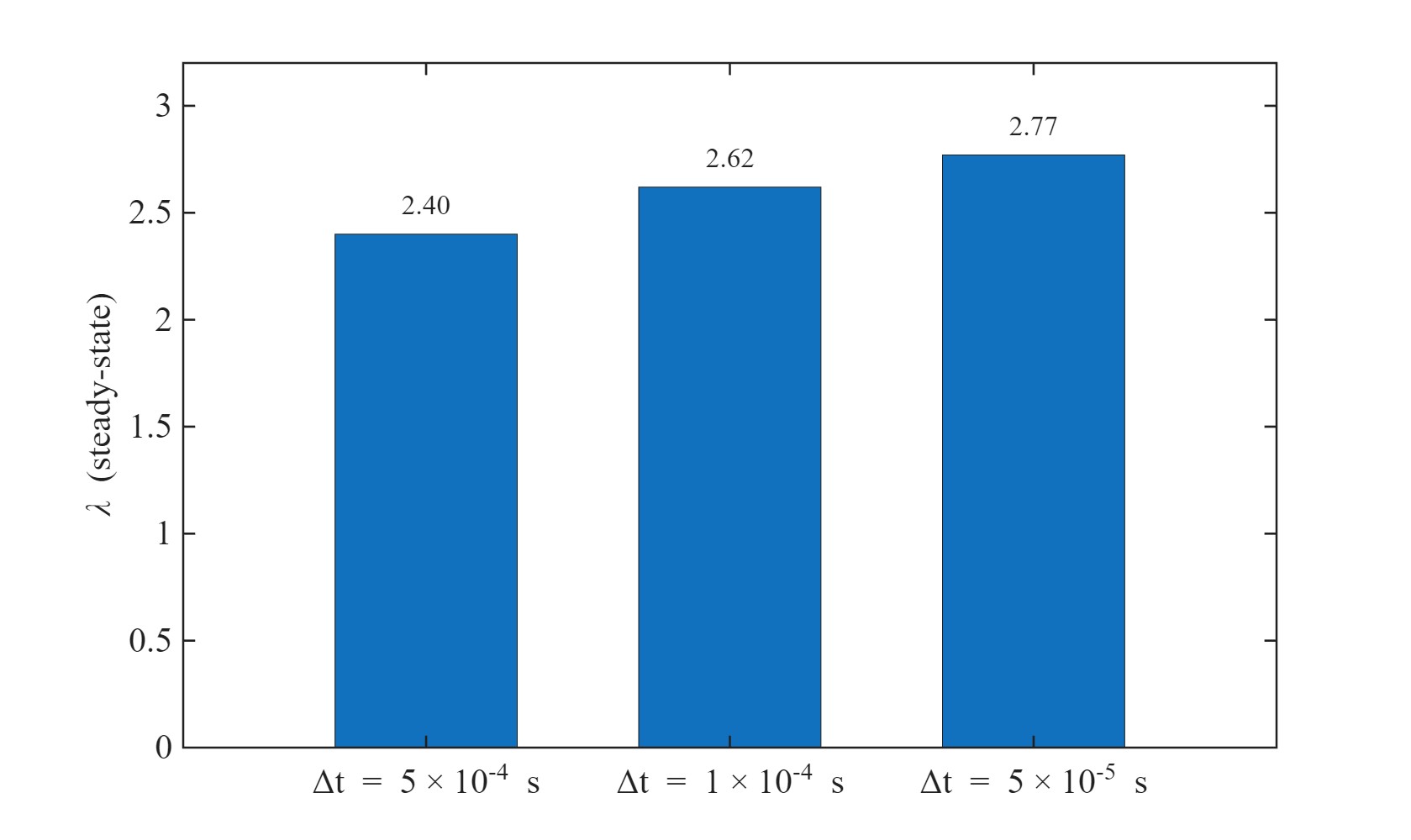}
  \caption{\edt{Comparison of the steady-state $\lambda$'s for three timestep sizes}}
  \label{fig:TimeInd}
\end{figure}

A time-step independence study is essential to ensure the numerical accuracy and reliability of transient simulations. In unsteady problems\edt{,} such as the self-starting process of VAWTs, the time-step size directly influences how well the solver can resolve rapid variations in aerodynamic forces, moments, and flow structures around the blades. In \edt{this work,} we select three \edt{timestep} sizes, \edt{including} $\Delta$$t_1=0.0005~\mbox{s}$, $\Delta$$t_2=0.0001~\mbox{s}$\edt{,} and $\Delta$$t_3=0.00005\mbox{s}$. We find that \edt{the tip-speed ratio of the VAWT for} $\Delta$$t_2$ and $\Delta$$t_3$ \edt{settles} at a $\lambda$ that does not have a significant difference\edt{,} as shown in Fig.~\ref{fig:TimeInd}, the former reaching steady-state $\lambda$ \edt{of} $2.77$ and the latter at $2.62$. Despite the small difference, $\Delta$$t_3$ is still selected for all subsequent simulations to provide greater temporal resolution and confidence in the results. 

\subsection{Validation}
\label{subsec:val}

\begin{figure}[H]
  \centering
  \includegraphics[width=0.9\linewidth]{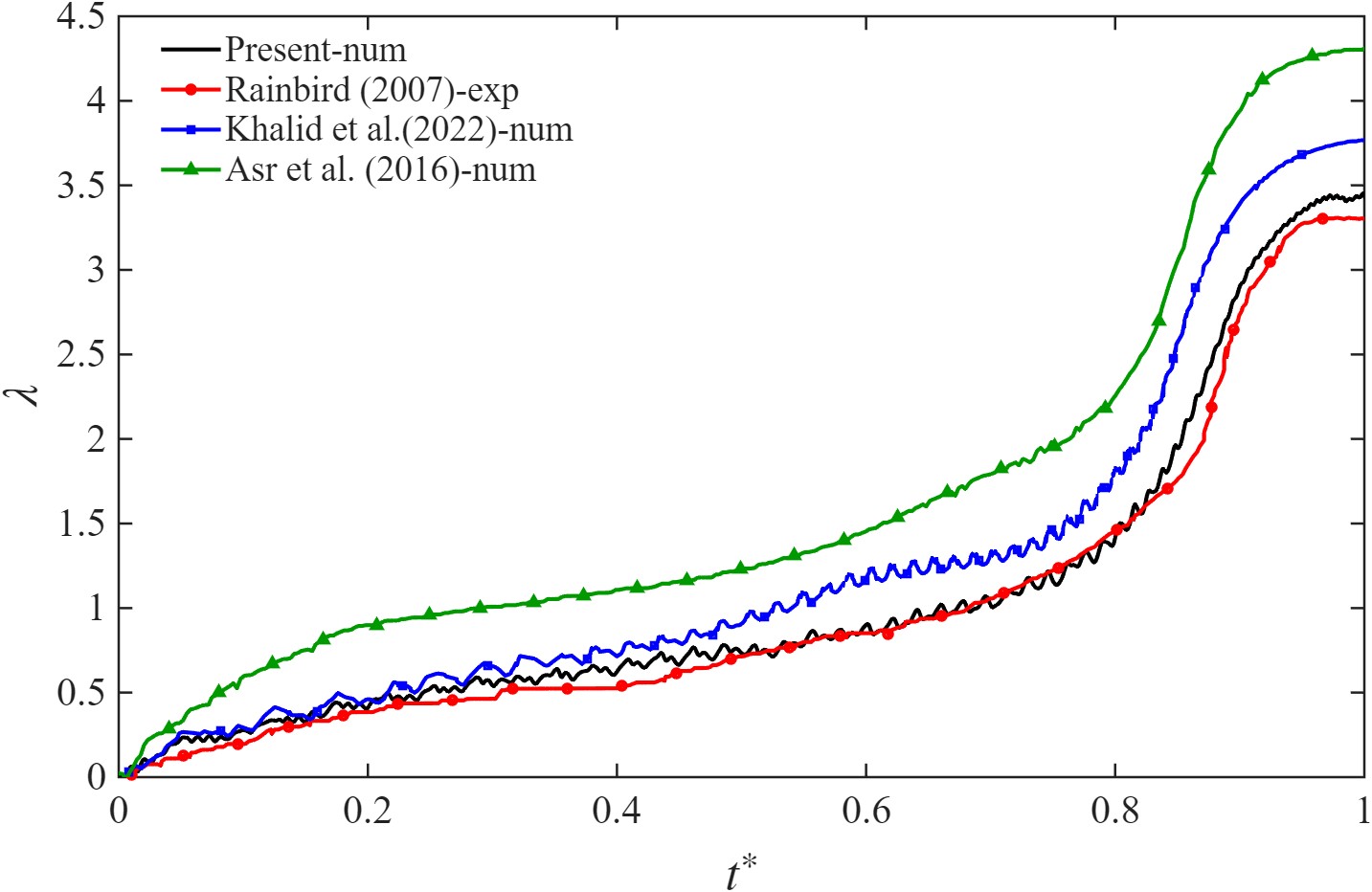}
  \caption{\edt{Comparison of $\lambda$ obtained from our computational simulation with experimental data of Rainbird \cite{rainbird2007aerodynamic} and numerical results from \fm{Khalid et al.} \cite{khalid2022self} and \fm{Asr et al.} \cite{asr2016study}}}
  \label{fig:val}
\end{figure}

\edt{Now,} we validate our \edt{computational framework} \edt{using} the experimental data \edt{provided by} \fm{Rainbird} \cite{rainbird2007aerodynamic} and the numerical results of \fm{Khalid et al.} \cite{khalid2022self} and \fm{Asr et al.} \cite{asr2016study}. \fm{The VAWT geometry used in these studies consists of a 3-bladed turbine with $c=83$~mm, $R=0.375$~m, and $J$ = 0.018~~\(\mathrm{kg\,m^2}\), subjected to a wind speed of $6~\mathrm{m/s}$.} As shown in Fig.~\ref{fig:val}, the four stages of the \edt{VAWT's} self‐starting process, described by \fm{Du et al.} \cite{du2019review}, are clearly captured. \edt{These stages include an} initial nearly-linear acceleration, a plateau with only a slow increase in $\lambda$, a rapid–acceleration phase, and finally convergence to a quasi–steady operating $\lambda$. The comparison indicates very good agreement \edt{of our presently obtained results} with the experimental \edt{data} while remaining consistent with \edt{previously reported} simulations. The experiment employs a finite–span rotor and includes resistive torques from the test rig and drivetrain, whereas our present two–dimensional CFD assumes an infinite span and a freely rotating turbine without structural or generator damping. Although such idealizations are known to be mildly over-predictive relative to experiments, the close correspondence of our curves with \fm{Rainbird} \cite{rainbird2007aerodynamic} and their alignment with \edt{the results from \fm{Khalid et al.} \cite{khalid2022self} and \fm{Asr et al.} \cite{asr2016study}, provide} confidence in \edt{our} \edt{computational} setup and supports the subsequent analysis of self–starting behavior \edt{of VAWTs} in this work.

\section{Results and Discussion}
\label{sec:resrults}

\edt{The results from our investigations are} organized around the two configuration sets\edt{, including the $\mbox{EC}$ set and the $\mbox{ES}$ set} introduced in Section~\ref{subsec:Geom}. Within each set, comparisons are made between the \edt{$3$-bladed} and \edt{$5$-bladed turbines} to isolate the effect of \edt{the number of blades} $N$. In the $\mbox{EC}$ set, the \edt{the turbine with $3$ blades} establish the reference trend in $\lambda_{\mathrm{steady}}$ at lower $\sigma$, while the \edt{$5$-bladed VAWT} isolate the effect of \edt{the} increasing $N$ at \edt{a} fixed $c$, which results in higher $\sigma$. In the $\edt{ES}$ set, the chord of \edt{a blade for the $5$-bladed turbine} is reduced to match the solidity of the corresponding \edt{$3$-bladed turbine}, isolating the effect of $N$ at \edt{a constant} $\sigma$. Together, these comparisons enable a systematic assessment of how $c$, $N$, and $\sigma$ influence the self-starting torque, transient acceleration, and $\lambda_{\mathrm{steady}}$ \edt{of VAWTs} at inflow velocities of $4~\mbox{m/s}$, $6~\mbox{m/s}$, and $8~\mbox{m/s}$.

\begin{figure}[H]
  \centering
  \includegraphics[width=1\linewidth]{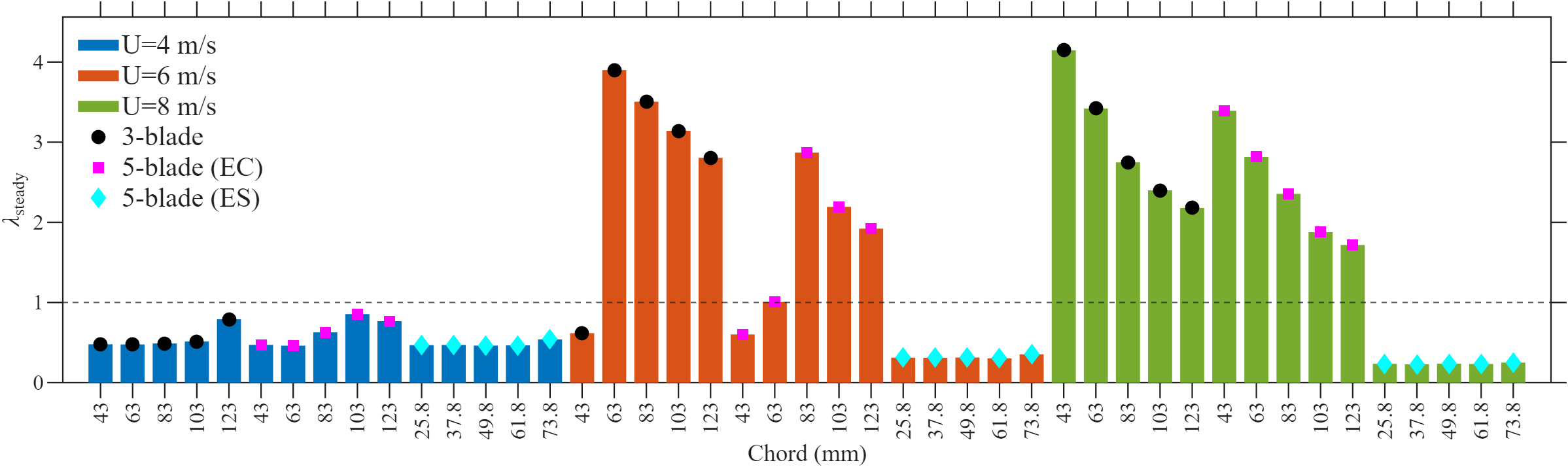}
  \caption{\fm{Bar chart summarizing $\lambda_{\text{steady}}$, versus $c$ for the EC and ES 3- and 5-bladed configurations at $U_{\infty}=4,\;6,\; \mbox{and}\ 8~\mathrm{m\,s^{-1}}$ with the dashed line representing $\lambda_{steady}=1$}}
  \label{fig:summ}
\end{figure}

\edt{Figure}~\ref{fig:summ} \edt{exhibits} $\lambda_{\mathrm{steady}}$ as a function of \edt{the} chord \edt{of a blade} for all configurations \edt{under the aforementioned flow conditions}. For the \edt{$3$-bladed} configurations, $\lambda_{\mathrm{steady}}$ increases \edt{significantly} with \edt{the} inflow velocity for all chord lengths. At \edt{the velocity of $4$~\mbox{m/s}}, all cases \edt{demonstrate $\lambda_{\mathrm{steady}} < 1$} and therefore do not self-start. At \edt{${U_\infty}=6~\mbox{m/s}$}, $\lambda_{\mathrm{steady}}$ increases sharply, reaching values above $3$ for intermediate chord lengths and decreasing slightly for \edt{the larger chord-lengths}. \edt{For ${U_\infty}=8~\mbox{m/s}$}, the \edt{$3$-bladed turbine attains} the highest $\lambda_{\mathrm{steady}}$, with \edt{its} clear monotonic \edt{decrement} as \edt{the} chord \edt{of a blade} increases from \edt{$43~\mbox{mm}$ to $123~\mbox{mm}$}. For the \edt{$5$-bladed turbines in the} $\mbox{EC}$ configurations, $\lambda_{\mathrm{steady}}$ is consistently lower than for the corresponding \edt{$3$-bladed counterparts} at the same chord \edt{of each blade} and \edt{the} inflow velocity. \edt{For ${U_\infty}=4~\mbox{m/s}$, $\lambda_{\mathrm{steady}}$ of all the} cases \edt{considered here remains} below unity and\edt{,} therefore\edt{,} do not self-start. At \edt{${U_\infty}=6~\mbox{m/s}$}, self-starting \edt{of turbines} occurs only for \edt{the} intermediate chords, and $\lambda_{\mathrm{steady}}$ remains below the \edt{corresponding values for the $3$-bladed VAWTs}. \edt{For ${U_\infty}=8~\mbox{m/s}$, the configuration in the $\mbox{EC}$ set} exhibit moderate $\lambda_{\mathrm{steady}}$ that decreases with \edt{an} increasing chord \edt{of a blade}. For the \edt{5-bladed} $\mbox{ES}$ configurations, $\lambda_{\mathrm{steady}}$ remains low for all inflow velocities and chord lengths. Even at \edt{${U_\infty}=8~\mbox{m/s}$}, $\lambda_{\mathrm{steady}}$ does not exceed approximately $0.4$, and none \edt{of the turbines in this set exhibits self starting behavior}. Overall, Fig.~\ref{fig:summ} shows that increasing $N$ at \edt{a} fixed $c$ reduces $\lambda_{\mathrm{steady}}$, while increasing $N$ at \edt{a} fixed $\sigma$ leads to a stronger suppression of self-starting and sustained rotation. The highest $\lambda_{\mathrm{steady}}$ occurs for the \edt{3-bladed} configurations at the lowest $\sigma$ and highest inflow velocity.

\begin{figure}[H]
  \centering
  \includegraphics[width=1\linewidth]{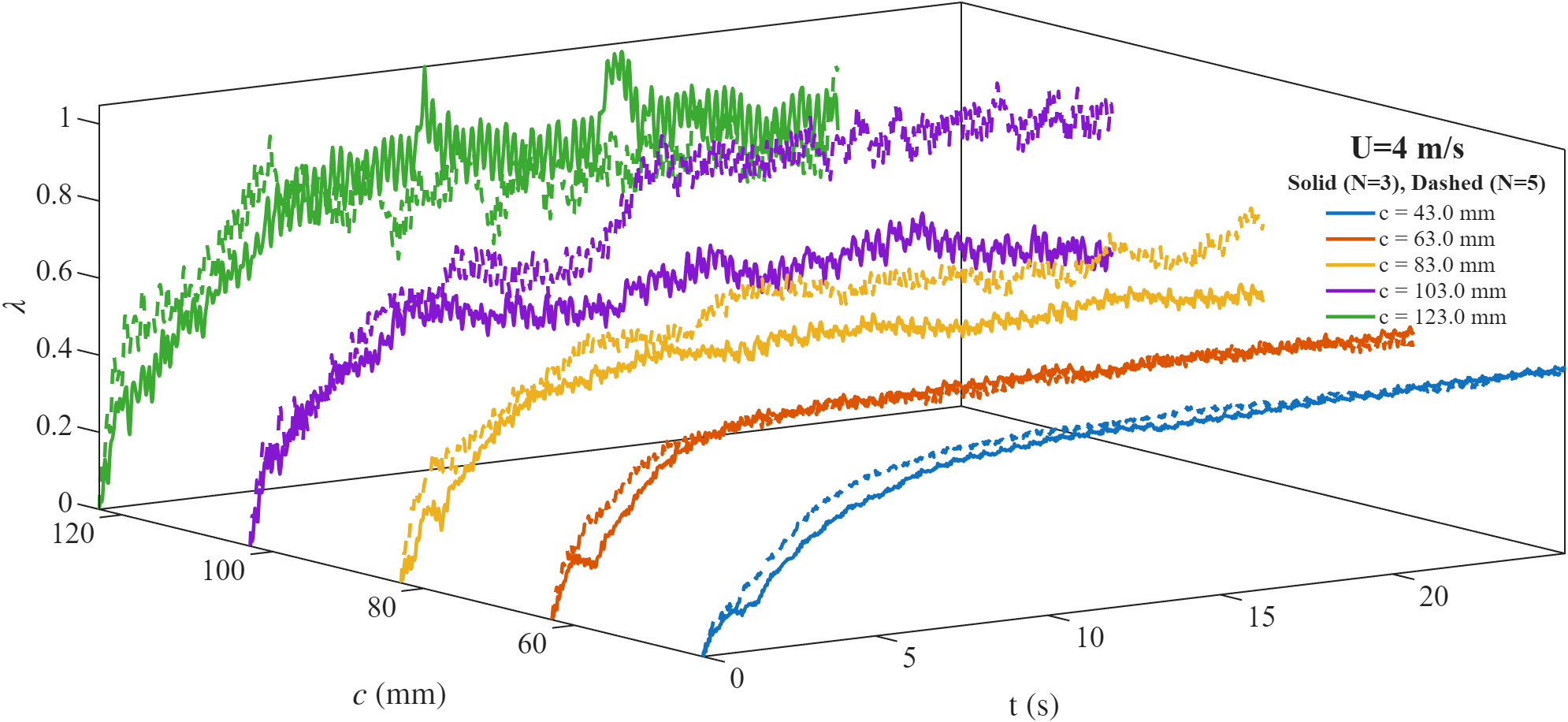}
  \caption{Comparative time history of $\lambda$ for the 3 and 5 blade turbines in the EC configuration at $U_\infty = 4~\mathrm{m/s}$.}
  \label{fig:EC_4ms}
\end{figure}
\begin{figure}[H]
  \centering
  \includegraphics[width=1\linewidth]{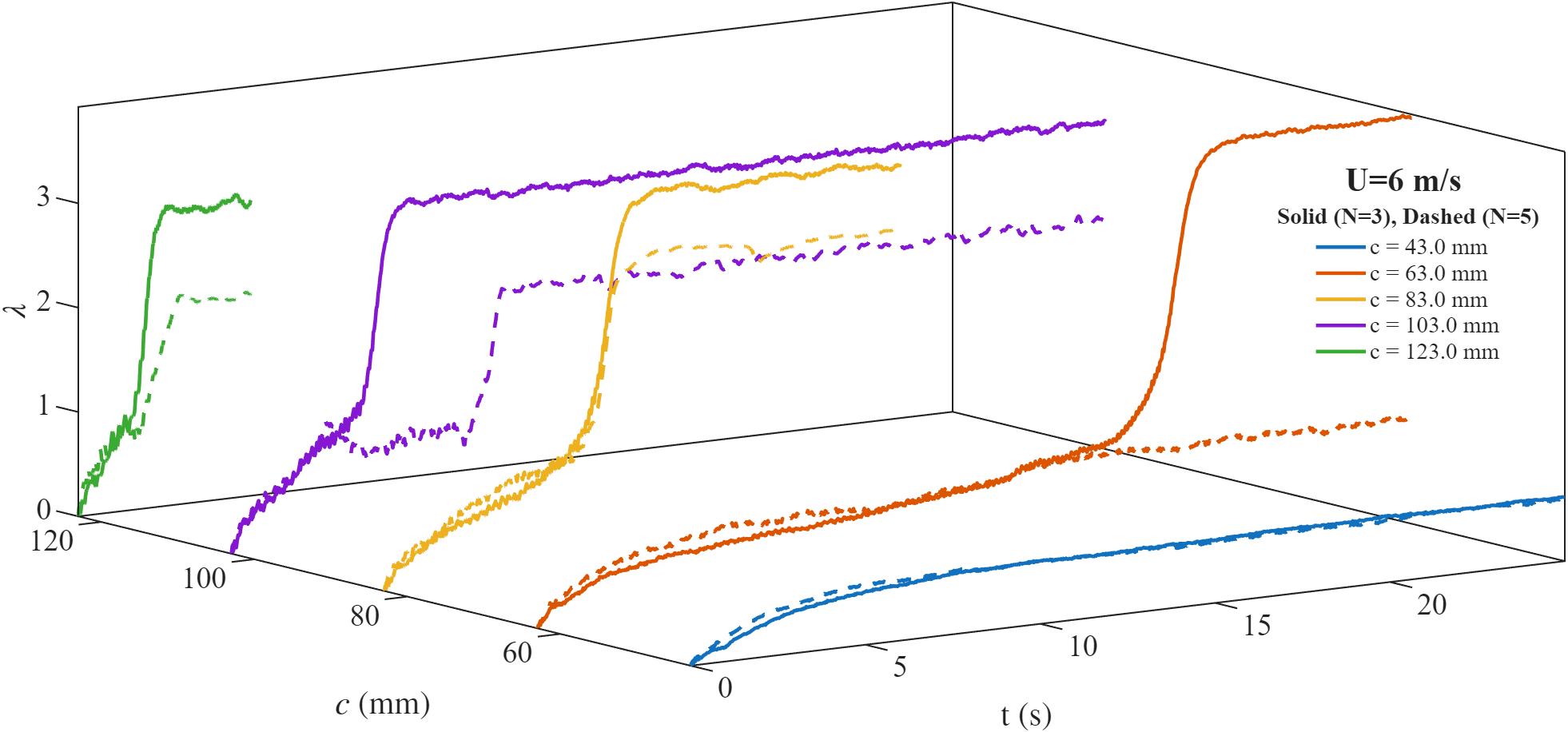}
  \caption{Comparative time history of $\lambda$ for the 3 and 5 blade turbines in the EC configuration at $U_\infty = 6~\mathrm{m/s}$.}
  \label{fig:EC_6ms}
\end{figure}
\begin{figure}[H]
  \centering
  \includegraphics[width=1\linewidth]{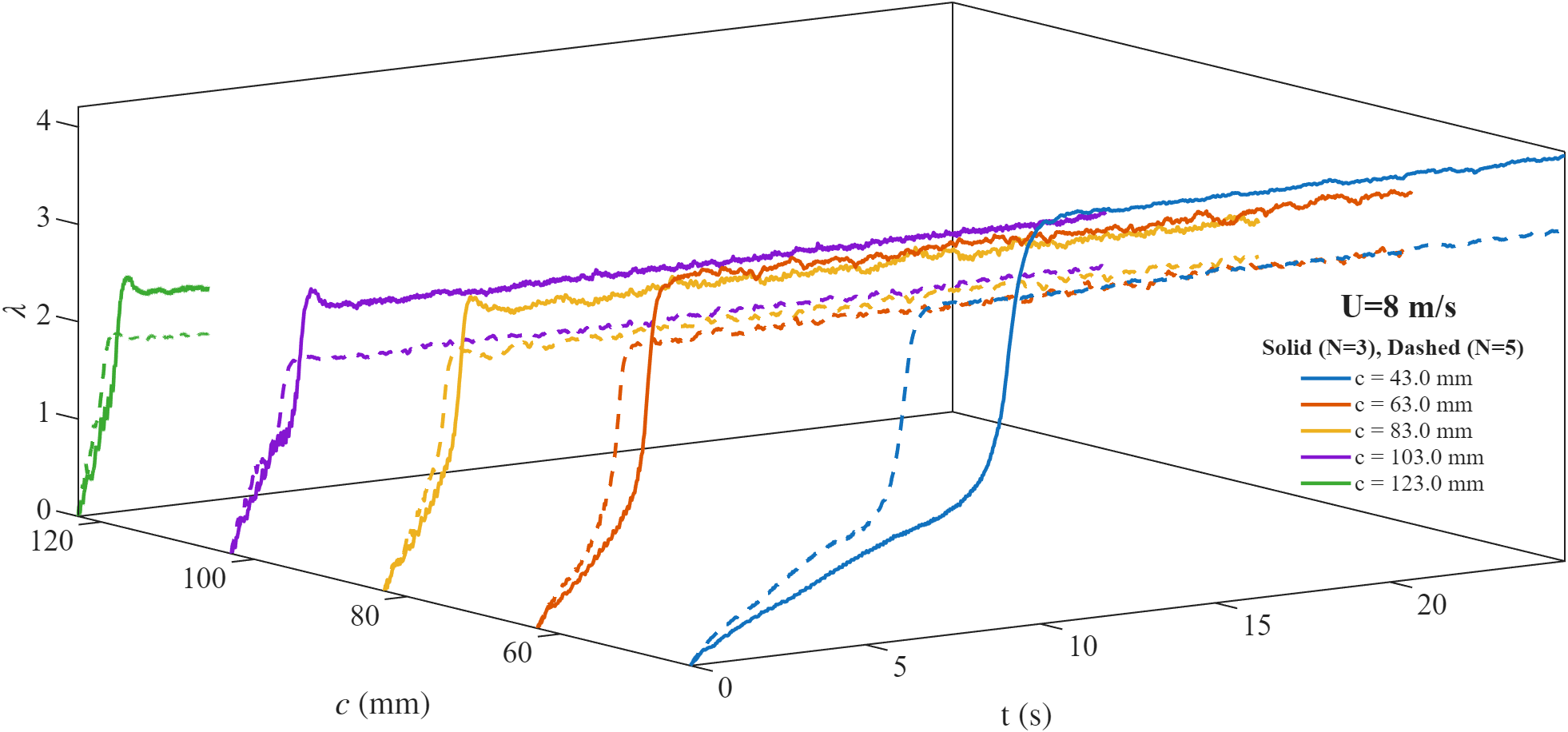}
  \caption{Comparative time history of $\lambda$ for the 3 and 5 blade turbines in the EC configuration at $U_\infty = 8~\mathrm{m/s}$.}
  \label{fig:EC_8ms}
\end{figure}
\begin{figure}[H]
  \centering
  \includegraphics[width=1\linewidth]{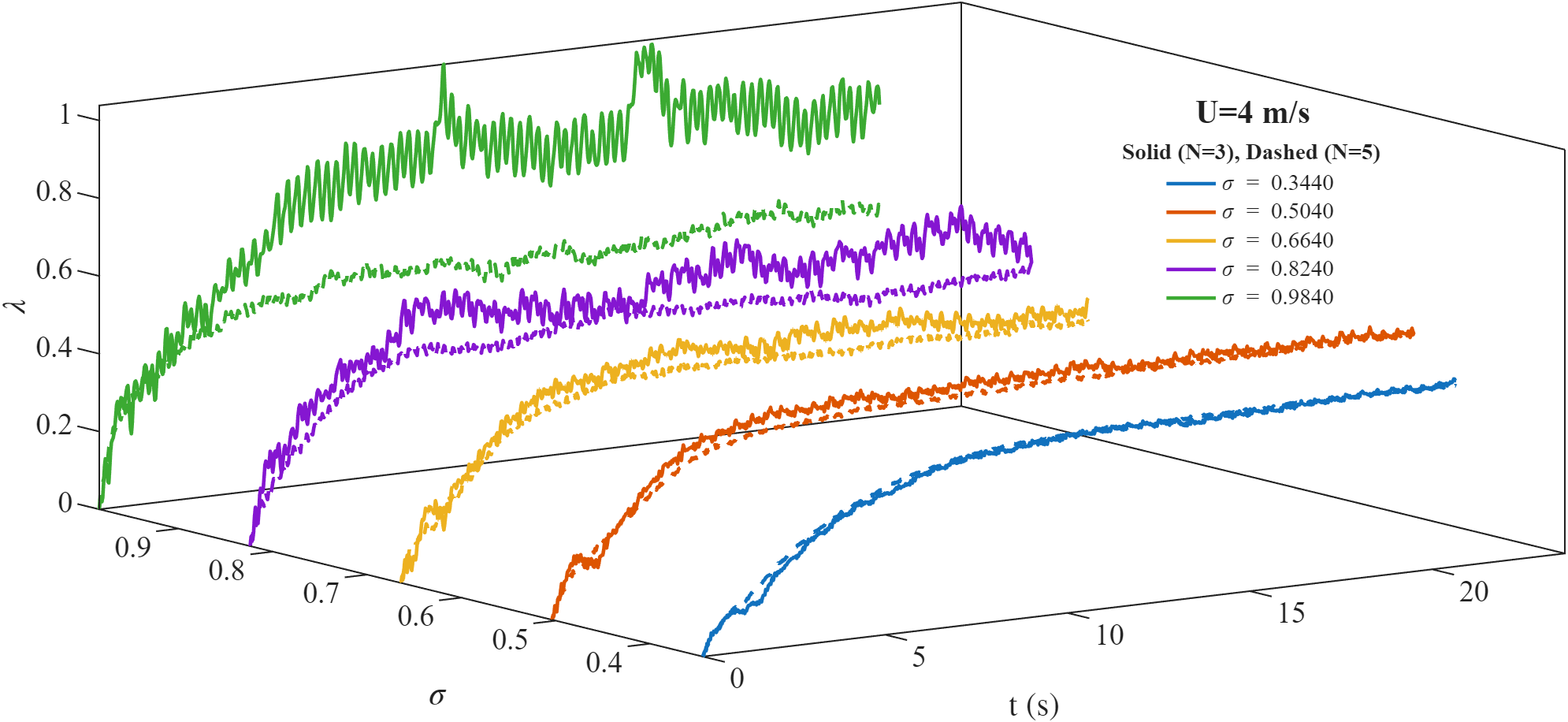}
  \caption{Comparative time history of $\lambda$ for the 3 and 5 blade turbines in the ES configuration at $U_\infty = 4~\mathrm{m/s}$.}
  \label{fig:ES_4ms}
\end{figure}
\begin{figure}[H]
  \centering
  \includegraphics[width=1\linewidth]{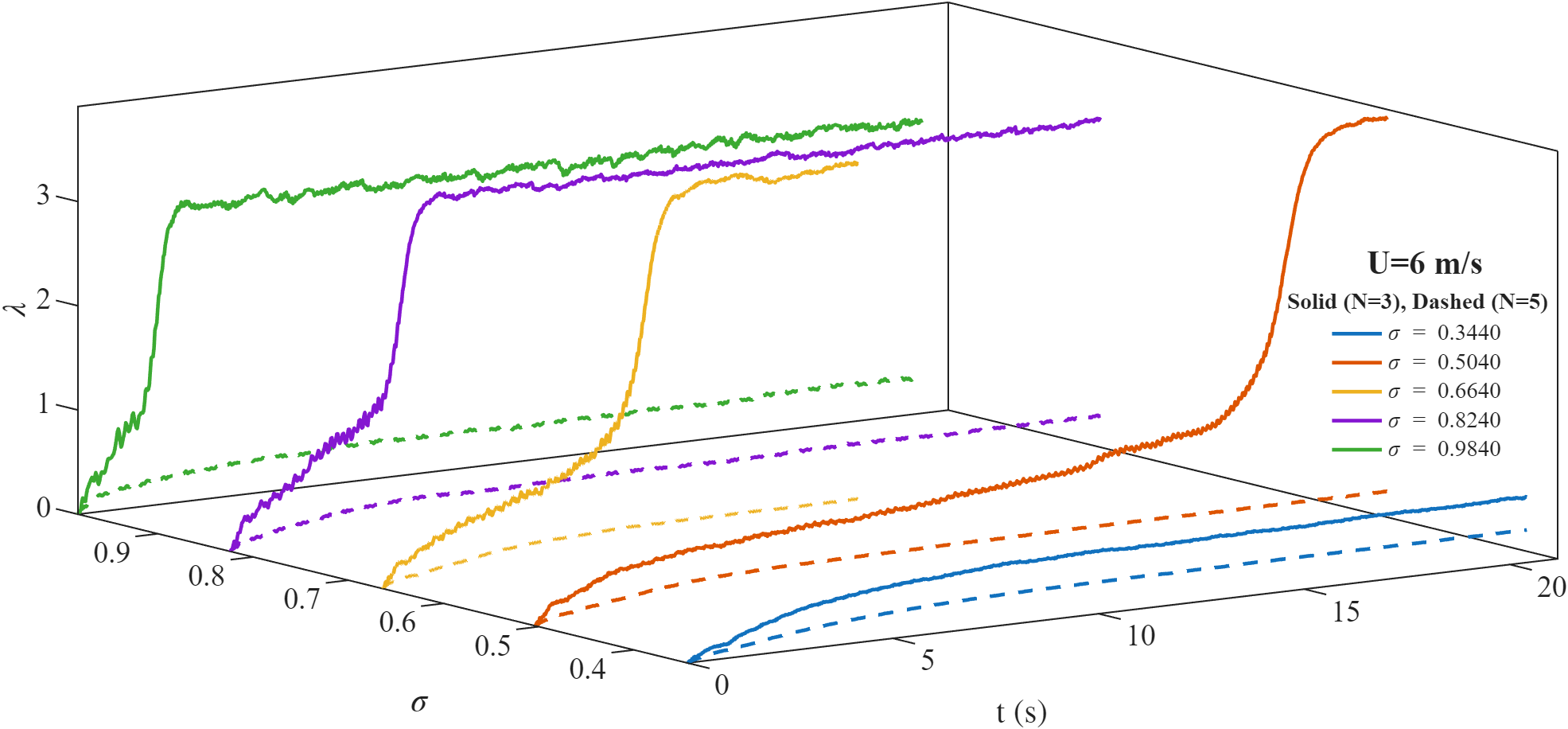}
  \caption{Comparative time history of $\lambda$ for the 3 and 5 blade turbines in the ES configuration at $U_\infty = 6~\mathrm{m/s}$.}
  \label{fig:ES_6ms}
\end{figure}
\begin{figure}[H]
  \centering
  \includegraphics[width=1\linewidth]{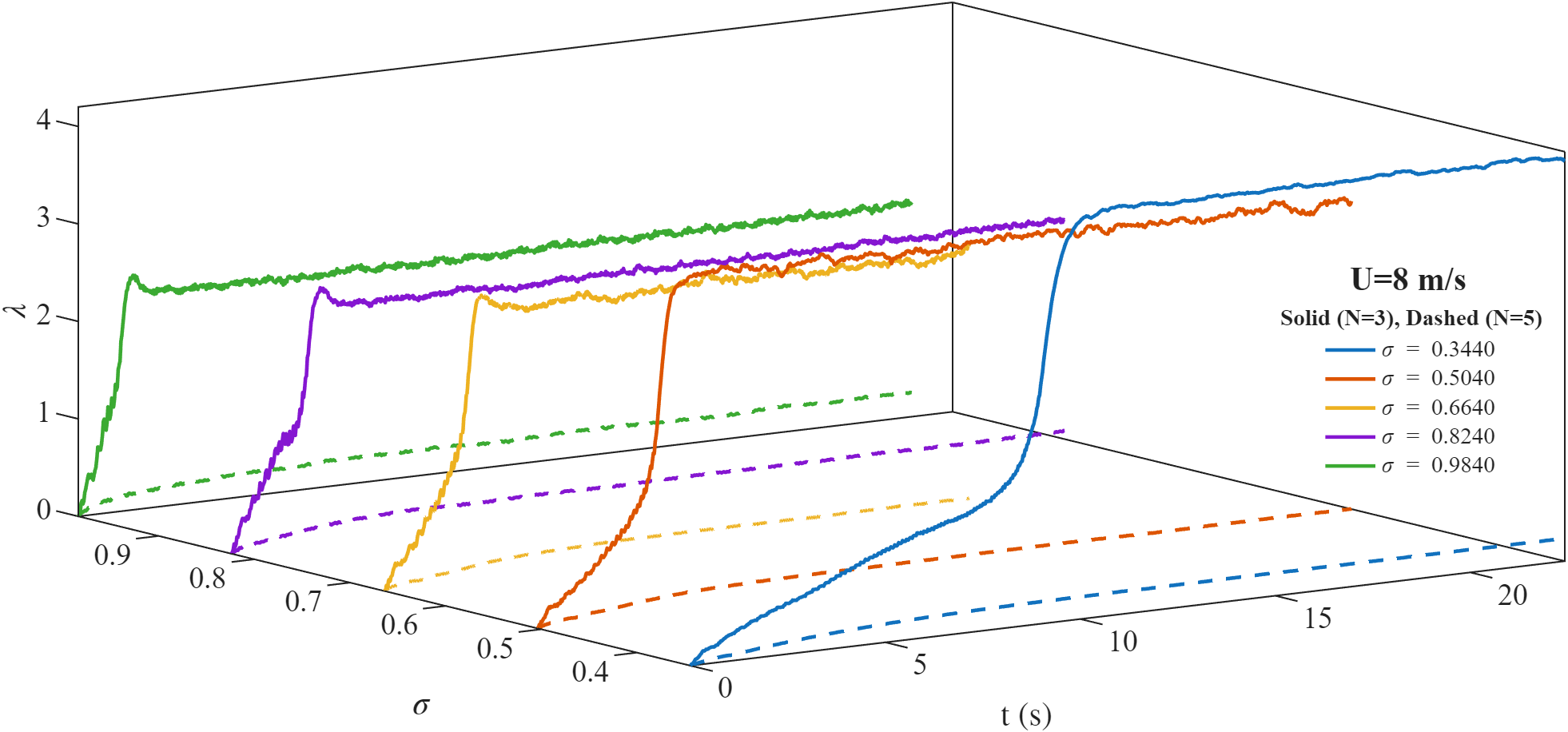}
  \caption{Comparative time history of $\lambda$ for the 3 and 5 blade turbines in the ES configuration at $U_\infty = 8~\mathrm{m/s}$.}
  \label{fig:ES_8ms}
\end{figure}

Figure~\ref{fig:summ} also reveals the existence of a critical chord length ($c_{\mathrm{crit}}$) for each configuration. \edt{It is clear that the only turbines with $c \geq c_{\mathrm{crit}}$ are able to self start and} overcome the dead band \edt{to} show self starting. \edt{Therefore,} the value of $c_{\mathrm{crit}}$ represents the minimum $\mbox{c}$ required for self starting \edt{of a VAWT} for a given $N$, $R$, and $J$. It also corresponds to the chord that maximizes $\lambda_{\mathrm{steady}}$ within each configuration family, indicating a trade off between startup capability and steady state performance\edt{. These trends are also observed from the temporal profiles of $\lambda$ presented} in \edt{Figs}.~\ref{fig:EC_4ms} to \ref{fig:ES_8ms}. \edt{Please note that the aim here is not to} determine $c_{\mathrm{crit}}$ precisely in this study, \edt{but providing a good estimates of the range of the geometric characteristics of a VAWT capable of self starting,} as summarized in Table~\ref{tab:ccrit_bounds}.


\begin{table}[h]
\centering
\caption{\edt{Ranges for} $c_{\mathrm{crit}}$ separating non-self-starting ($\lambda_{\mathrm{steady}}<1$) and self-starting ($\lambda_{\mathrm{steady}}>1$) regimes}
\label{tab:ccrit_bounds}

\renewcommand{\arraystretch}{1.15}
\setlength{\tabcolsep}{10pt}

\begin{tabular}{lccc}
\toprule
\textbf{$U_\infty$ (m/s)} & \textbf{3 blades} & \textbf{5 blades (EC)} & \textbf{5 blades (ES)} \\
\midrule
4 & $c_{\mathrm{crit}} > 123~\mathrm{mm}$ & \textbf{$103 < c_{\mathrm{crit}} < 123~\mathrm{mm}$} & $c_{\mathrm{crit}} > 73.8~\mathrm{mm}$ \\
6 & $43 < c_{\mathrm{crit}} < 63~\mathrm{mm}$ & $63 < c_{\mathrm{crit}} < 83~\mathrm{mm}$ & $c_{\mathrm{crit}} > 73.8~\mathrm{mm}$ \\
8 & $c_{\mathrm{crit}} < 43~\mathrm{mm}$ & $c_{\mathrm{crit}} < 43~\mathrm{mm}$ & $c_{\mathrm{crit}} > 73.8~\mathrm{mm}$ \\
\bottomrule
\end{tabular}
\end{table}

\edt{Next, Figs.}~\ref{fig:EC_4ms}\edt{, \ref{fig:EC_6ms}, and} \ref{fig:EC_8ms} present the time \edt{histories} of $\lambda$ for the \edt{$3$-bladed} and \edt{$5$-bladed} turbines in the $\mbox{EC}$ configuration at $U_\infty = 4~\mathrm{m/s}$, $6~\mathrm{m/s}$, and $8~\mathrm{m/s}$, respectively. \edt{For} all \edt{the} three inflow velocities, the \edt{$5$-bladed turbine} accelerates more rapidly from rest and reaches its steady state earlier than the \edt{\fm{$3$}-bladed turbine}, indicating a shorter startup time. However, $\lambda_{\mathrm{steady}}$ attained by the \edt{$5$-bladed turbine} is consistently lower than that of the \edt{$5$-bladed turbine}. \edt{It} demonstrates that increasing the number of blades enhances the initial aerodynamic torque but penalizes the long term rotational performance. \fm{Across the EC and ES configurations, the present results show that the effect of $N$ on self-starting is strongly conditioned by how the geometry is modified. In the EC set (same $c$), increasing $N$ from $3$ to $5$ generally promotes a faster initial rise in $\lambda$ and a shorter startup period, particularly at $U_{\infty}=6$ and $8~\mathrm{m/s}$, but it also leads to a lower $\lambda_{\mathrm{steady}}$ after startup. In contrast, in the ES set (same $\sigma$), the reduction in $c$ required for the $5$-bladed turbine suppresses blade-level aerodynamic loading, and the $5$-bladed cases remain trapped in the dead-band (non-self-starting) over the range considered, even when the corresponding $3$-bladed cases self-start. These observations indicate that increasing \edt{the number of blades} alone does not guarantee improved self-starting. Rather, its \edt{influence} depends on the associated \edt{scaling of} $c$ and the resulting unsteady aerodynamic response. \edt{It} is broadly consistent with \fm{Sun et al.} \cite{sun2021rotation} in the sense that effects of $N$ are wind-speed dependent and can influence startup time but differs in the final outcome. It is because \fm{Sun et al.} \cite{sun2021rotation} considered turbines with an offsetting pitching angle (which can delay or suppress adverse vortex development), whereas the present work isolates fixed-pitch geometric effects in freely accelerating VAWTs. As a result, the present simulations highlight a startup--performance trade-off: a higher blade count may assist early acceleration under equal-$c$ conditions, yet it can reduce the attainable steady operating state and, under equal-$\sigma$ $chord$ reduction, inhibit self-starting altogether.} 


\edt{Figures}.~\ref{fig:ES_4ms}\edt{, \ref{fig:ES_6ms}, and} \ref{fig:ES_8ms} show the corresponding time histories for \edt{of the turbines, belonging to} the $\mbox{ES}$ configuration at $U_\infty = 4~\mathrm{m/s}$, $6~\mathrm{m/s}$, and $8~\mathrm{m/s}$\edt{, respectively.} The \edt{$5$-bladed turbine} does not achieve self starting for any of the inflow velocities considered \edt{here}. \edt{It is found that} $\lambda$ remains low and increases only gradually with time, indicating that the turbine remains trapped in the dead band and does not reach a sustained rotational state. \edt{Contrarily}, the \edt{$3$-bladed} turbine is able to self start and reach a steady $\lambda$. \edt{It} demonstrates that, for the $\mbox{ES}$ family \edt{of VAWTs}, increasing the blade number inhibits self starting rather than promoting it. 

\subsection{Dynamic Stall and Blade--Vortex Interaction}

In this \edt{sub}section, the development of vortical structures, blade--vortex \edt{interactions} (BVI), and their influence on both the startup process and the steady-state behavior are examined. Particular attention is paid to the role of dynamic stall in shaping the unsteady aerodynamic loads experienced by the blades and \edt{production of the overall torque by the turbines}.

Dynamic stall is generally \edt{considered} to occur on an airfoil or \edt{a} lifting surface when it is subjected to time-dependent \edt{oscillatory} motion\edt{,} such that the effective angle\edt{-}of\edt{-}attack \edt{(}$\alpha_{\mathrm{eff}}$\edt{)}, exceeds the corresponding static stall angle, \edt{(}$\alpha_{\mathrm{ss}}$\edt{)} \edt{\citep{mccroskey1976dynamic,mcalister1982experimental}}. This phenomenon is inherently unsteady and depends not only on the instantaneous value of $\alpha_{\mathrm{eff}}$, but also on its rate of change and on the kinematics of the \edt{surface}. Consequently, the distinction between static and dynamic stall is closely related to the rate of change of angle\edt{-}of\edt{-}attack \edt{(}$\dot{\alpha}$\edt{)}. However, no universal quantitative definition exists that uniquely separates static and dynamic stall based on $\dot{\alpha}$ alone \edt{\citep{le2021dynamics}}. Instead, dynamic stall is commonly characterized using non-dimensional parameters that reflect the level of flow unsteadiness. One such parameter that measures the degree of unsteadiness in the flow is the reduced frequency \fm{($k$)} \edt{\citep{leishman2006principles}, which is} defined \fm{with respect to $\theta$} as:

\begin{equation}
k_{\theta} = \frac{\omega c}{2U_{\mathrm{eff}}},
\end{equation}

This parameter was originally introduced for \edt{aerodynamics of rotors of helicopters} but is also applicable to VAWTs, where it provides a measure of the unsteadiness associated with the azimuthal rate of change of the blade. {More directly relevant to dynamic stall in VAWTs is the reduced frequency associated with pitching motion} \fm{\cite{le2022dynamic}} \edt{which is} defined \edt{below}:

\edt{
\begin{equation}
k_{\alpha} = \frac{{\dot{\alpha}_{eff}}{c}}{2U_{\mathrm{eff}}},
\end{equation}
}

\fm{As noted by Leishman \cite{leishman2006principles}, dynamic stall may arise once $k$ enters the unsteady regime. When $k = 0$, the flow is steady, implying no lag in the response of the surrounding fluid to the blade kinematics. For $0 \leq k < 0.05$, the flow may be regarded as quasi-steady, and unsteady effects are generally weak and often negligible. In contrast, for $k \geq 0.05$, the flow becomes increasingly unsteady, and dynamic stall effects may become significant.}

\edt{From a physical standpoint}, this unsteadiness can be interpreted as a delay in the response of the surrounding flow field to the blade\edt{'s} motion. When \edt{its} kinematics vary sufficiently rapidly, the flow is unable to adjust instantaneously to the imposed motion. {This delayed response causes flow reversal within the boundary layer, initiating near the trailing edge and progressively moving upstream} once \edt{$\alpha_{\mathrm{eff}}$} exceeds $\alpha_{\mathrm{ss}}$. As the upstream influence of this reversed flow reaches the leading edge, the separated shear layer rolls up into a concentrated leading-edge vortex, commonly referred to as the dynamic stall vortex ($\mbox{DSV}$). As the $\mbox{DSV}$ convects downstream over the suction surface, it induces a transient increase in lift\edt{. Its} continued convection and interaction with oppositely signed vorticity ultimately destabilize the vortex, leading to its detachment and a sharp drop in the moment coefficient. \edt{Afterwards,} the flow progresses toward a fully separated state until reattachment occurs at \edt{a lower $\alpha$}.

In the context of VAWTs, the \edt{recent} experimental work of \fm{Le Fouest and Mulleners} \citep{le2022dynamic} is particularly relevant, as they quantified the \edt{delay in the} aerodynamic response  using a single blade subjected to prescribed rotation and distinguished between deep stall, light stall, and no-stall regimes. Deep stall corresponded to cases in which the dynamic stall vortex had sufficient time to develop fully before detaching, leading to strong unsteady loads, whereas light stall occurred when the blade reached its maximum $\alpha_{\mathrm{eff}}$ before the vortex \edt{was} fully formed, resulting in \edt{a} weaker unsteady loading. The no-stall regime was associated with conditions under which $\alpha_{\mathrm{eff}}$ did not exceed $\alpha_{\mathrm{ss}}$, which in their study corresponded to \edt{operations at large $\lambda$'s}. \edt{This approach of quantifying the} dynamic\edt{-}stall regimes is not directly applicable in \edt{our} present work. \edt{The experiments done by \fm{Le Fouest and Mulleners} \cite{le2022dynamic} were based on a single-blade geometry with predefined rotations, which is rather a simple scenario. Our present simulations} consider the collective passive aerodynamic response of \edt{the VAWTs composed of} multiple blades. \edt{Consequently}, it is \edt{difficult} to clearly identify the instant at which an individual blade exceeds its static stall angle\edt{,} or the corresponding instant at which the pitching moment coefficient reaches its maximum \edt{value}, which are required to compute the \edt{delay in the response}. Instead, \edt{we examine} the unsteady behaviour using the \edt{temporal histories} of instantaneous $\theta$, $\alpha_{eff}$, $\lambda$, $k_{\alpha}$ and $k_{\theta}$, which provide qualitative and comparative \edt{insights for} the presence and nature of dynamic stall under different operating conditions.

\begin{figure}
  \centering
  \includegraphics[width=1.0\linewidth]{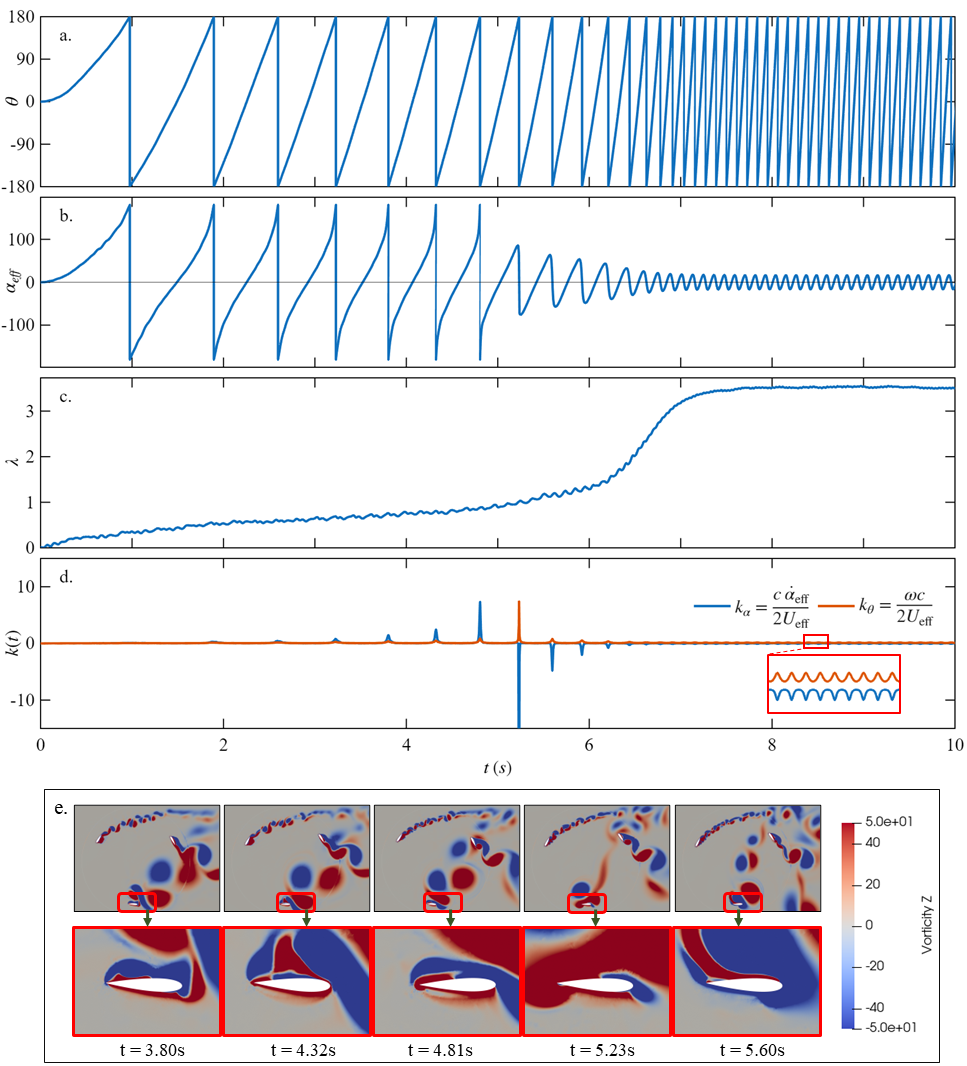}
  \caption{Single-blade kinematic time histories for a self-starting case with $N=3$, $c=83$~mm, and $U_{\infty}=6$~m/s. Shown are (a) azimuthal position $\theta(t)$, (b) effective angle of attack $\alpha_{\mathrm{eff}}(t)$, (c) tip-speed ratio $\lambda(t)$, (d) reduced frequencies $k_{\alpha}(t)$ and $k_{\theta}(t)$, and (e) Vorticity Fields at specified instants \fm{figure changed to add the zoomed in section}}
  \label{fig:kalpha_case1}
\end{figure}

To investigate the role of dynamic stall in the self-starting process, a representative \edt{$3$-bladed VAWT} exhibiting self-starting behavior is selected for the detailed analysis. The instantaneous evolution of the kinematic quantities introduced earlier, together with the associated vorticity fields for the case with $U_{\infty}=6$~m/s and $c=83$~mm, is presented in \edt{Fig.~\ref{fig:kalpha_case1}}. The corresponding self-starting \edt{$5$-bladed turbines} exhibit qualitatively similar kinematic and flow features, therefore, only the \edt{$3$-bladed configuration} is discussed here for brevity, without loss of generality. From \edt{Fig.~\ref{fig:kalpha_case1}a}, the evolution of the azimuthal position illustrates how the rotational speed increases with time, while \edt{Fig.~\ref{fig:kalpha_case1}c} shows that\edt{, in the initial phase of the process,} the turbine operates at very low values of $\lambda$. \edt{Figure~\ref{fig:kalpha_case1}b presents evidently} that $\alpha_{\mathrm{eff}}$ experiences very large excursions at the beginning of the rotation, well beyond typical static stall angles, which \edt{were} reported in literature to lie approximately between $13^{\circ}$ and $20^{\circ}$ for NACA0018 depending on Reynolds number \cite{timmer2008two, stangfeld2015unsteady, damiola2023influence, le2021dynamics}. However, large excursions of $\alpha_{\mathrm{eff}}$ alone are not sufficient to infer the onset of dynamic stall, since the existence\edt{, strength, and influence of} dynamic stall depend on the unsteady time scale of a blade's motion. \edt{The reduced frequency provides a measure to quantify these effects.} The unsteadiness in the flow becomes more pronounced at approximately $t \approx 1.8$~s, where a noticeable change in the reduced frequencies is observed. At this time instant, a small spike appears in \edt{Fig.~\ref{fig:kalpha_case1}d}, which coincides with the blade reaching $\theta \approx 180^{\circ}$. Similar spikes are observed in successive cycles at the same azimuthal position, with gradually increasing magnitude. At $t \approx 4.81$~s, a sharp increase in $k_{\alpha}$ is observed, indicating a very rapid change in $\alpha_{\mathrm{eff}}$ over a short time interval. A similar but comparatively smaller increase is also observed in $k_{\theta}$. As shown in \edt{Fig.~\ref{fig:kalpha_case1}d}, both reduced frequencies attain their maximum excursions at $t \approx 5.2~\mathrm{s}$. At this instant, $k_{\theta}$ exhibits a sharp positive spike\edt{, whereas} $k_{\alpha}$ undergoes a very large negative spike, indicating an abrupt change in the blade kinematics \edt{with} the blade \edt{passing} through $\theta \approx 180^{\circ}$\edt{,} where $U_{\mathrm{eff}}$ becomes small. Following this event, the magnitudes of both reduced frequencies decrease rapidly and only small fluctuations remain, while the turbine subsequently accelerates\edt{,} and $\lambda$ increases toward its steady value, as shown in \edt{Fig.~\ref{fig:kalpha_case1}c}.

Following the experimental observations of \fm{Hill et al.} \edt{\cite{hill2009darrieus} and \fm{Du et al.} \cite{du2019review}}, the self-starting process in the present case can be segmented into four distinct phases. The initial startup phase occurs for $0 < t < 1.9$~s, where $t = 1.9$~s corresponds to the time at which the first small surge in the reduced frequencies is observed, while only a nearly linear increase in $\lambda$ is present. The plateau stage, commonly referred to as the dead-band, extends from $1.9 < t < 5.2$~s. The time $t = 5.2$~s corresponds to the instant at which the largest positive peaks in both reduced frequencies are observed. During this interval, $\lambda$ remains largely unchanged, as shown in \edt{Fig.~\ref{fig:kalpha_case1}c}, indicating that the turbine experiences only a very gradual increase in \edt{its} angular velocity despite the presence of intermittent unsteady events. \edt{It} is followed by a rapid acceleration phase for $5.2 < t < 6.62$~s, during which the turbine undergoes runaway acceleration and $\lambda$ increases sharply. Finally, a steady operating phase is reached for $t > 6.62$~s. The steady operating phase is defined here as the regime in which $\alpha_{\mathrm{eff}}(t)$ \edt{exhibits a} sinusoidal \edt{profile with the rotational cycle of the turbine,} and both reduced frequencies exhibit a stable, \fm{quasi-}{periodic} pattern \fm{as shown in the magnified inset of Fig.~\ref{fig:kalpha_case1}d.}

\edt{Now, Fig.~\ref{fig:kalpha_case1}e} presents the vorticity fields at the instants corresponding to the spikes observed in the reduced frequencies. At $\theta = 180^{\circ}$, the \fm{suction side of the}blade transitions from the inward half of the trajectory to the outward half, a position at which the interaction between oppositely signed vorticity shear layers becomes particularly pronounced. During the first half of the rotation ($\theta = 0^{\circ}$ to $180^{\circ}$), the \edt{blade's angle-of-attack exceeds} $\alpha_{\mathrm{ss}}$, initiating boundary-layer flow reversal that develops from the trailing edge and progressively advances upstream. This reverse flow displaces the positive-vorticity shear layer associated with the inward side of the blade. By the time the blade reaches $\theta = 180^{\circ}$ at $t$ = 3.8s \fm{(the selected instant at which a second peak in $k_\theta$ and $k_\alpha$ is observed)} the flow is largely separated, consistent with the strong \edt{spikes in the reduced frequency} observed at this azimuthal position \edt{(see Fig.\ref{fig:kalpha_case1}d)}. As the turbine continues to accelerate\edt{,} and \edt{its} $\omega$ increases from cycle to cycle, the extent of boundary-layer separation is progressively reduced for \fm{$t$ > 5.23s as seen in Fig.~\ref{fig:kalpha_case1}e}. \edt{The contour plots of vorticity in Fig.~\ref{fig:kalpha_case1}e clearly shows that} the displaced positive-vorticity shear layer is observed to recover and migrate back toward the blade on the inward half of the trajectory in successive cycles. This recovery coincides with the growth and subsequent decay of the reduced-frequency spikes, indicating a gradual stabilisation of the near-blade flow as $\lambda$ increases.

\begin{figure}[h!]
  \centering
  \includegraphics[width=1\linewidth]{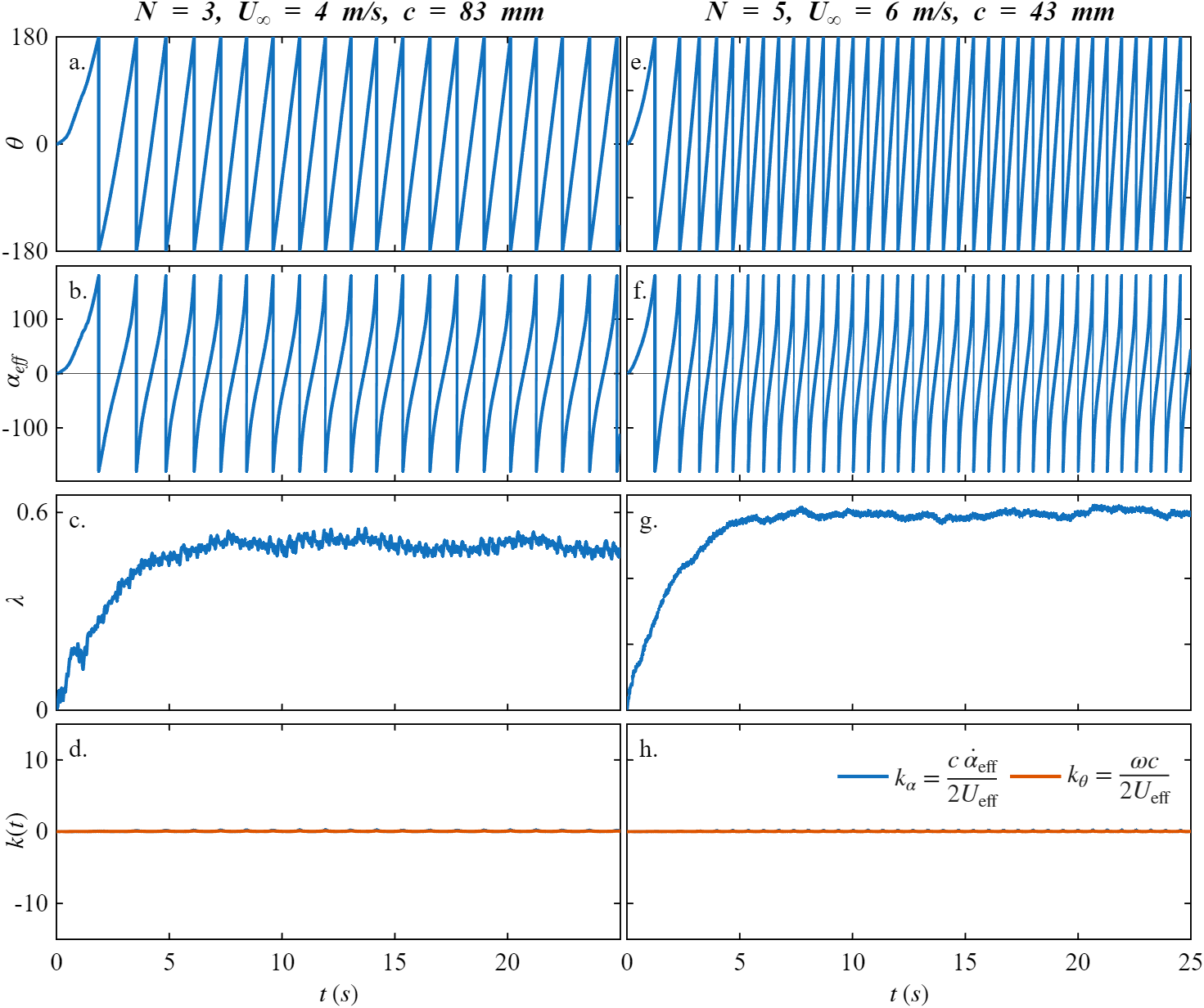}
  \caption{Single-blade kinematic time histories for two non-self-starting cases. Left column: $N=3$, $c=83$~mm, $U_{\infty}=4$~m/s. Right column: $N=5$, $c=43$~mm, $U_{\infty}=6$~m/s. Shown are (a,e) azimuthal position $\theta(t)$, (b,f) effective angle of attack $\alpha_{\mathrm{eff}}(t)$, (c,g) tip-speed ratio $\lambda(t)$, and (d,h) reduced frequencies $k_{\alpha}(t)$ and $k_{\theta}(t)$ }
  \label{fig:nonself}
\end{figure}

For the \edt{turbines with $3$ and $5$ blades} that do not self-start, the \edt{temporal} histories of $\theta$, $\alpha_{\mathrm{eff}}$, $\lambda$, and the reduced frequencies are shown in \edt{Fig.~\ref{fig:nonself}}. \edt{Both types of} turbines remain trapped within the dead-band regime, with $\lambda$ remaining below unity \edt{during the whole time}. Although brief excursions of $k_{\alpha}$ and $k_{\theta}$ above the quasi-steady threshold are observed, with peak values reaching approximately $0.1 - 0.2$, these events are intermittent and do not persist over consecutive cycles. Importantly, they are not accompanied by any sustained increase in $\lambda$, indicating that these localized unsteady events are insufficient to reorganize the near-blade flow in a manner that produces net positive torque and runaway acceleration. \edt{It is important to note that all} simulations \edt{in our present work are run for} for a sufficiently long physical time to ensure that the observed behaviour is not a transient artifact of the initial \edt{conditions. Particularly,} the non-self-starting cases \edt{are} run for more than $25~\mathrm{s}$ of physical time, consistent with startup durations commonly adopted in the literature \edt{\citep{celik2022design}}, and no delayed transition to self-starting \edt{is} observed. Over this duration, the flow \edt{remains} strongly separated for most of the cycle, with only limited intervals \edt{when $\alpha_{\mathrm{eff}} < \alpha_{\mathrm{ss}}$}. \edt{This observation confirms} that the absence of self-starting in these cases is a persistent dynamical outcome rather than a delayed transient response.


\edt{Further analysis of temporal profiles of the reduced frequency\fm{ies}} \fm{in Fig.~\ref{fig:kalpha_case1}d and boundary layer transition in Fig.~\ref{fig:kalpha_case1}e} reveal the physical origin of the two-stage sequence preceding self-starting. In the first stage \fm{at $t \approx 4.8\,\mathrm{s}$ (see Fig.~\ref{fig:kalpha_case1}e}), a pronounced peak appears in $k_{\alpha}$ alone. Since $k_{\alpha}$ scales with $\dot{\alpha}_{\mathrm{eff}}/U_{\mathrm{eff}}$, this initial peak is primarily driven by a rapid change in $\alpha_{\mathrm{eff}}$ over a short time interval, indicating a sudden reorientation of the relative inflow perceived by the blade. \edt{It} corresponds to a kinematic event in which the blade transitions rapidly between flow states, producing a large $\dot{\alpha}_{\mathrm{eff}}$ even though the azimuthal motion remains comparatively slow. In the subsequent cycle \fm{at $t \approx 5.23\,\mathrm{s}$, as shown in Fig.~\ref{fig:kalpha_case1}e }, a second peak occurs in $k_{\alpha}$ now accompanied by a concurrent peak in $k_{\theta}$. This second event reflects the combined effect of a rapid change in $\alpha_{\mathrm{eff}}$ and a strong kinematic amplification associated with the azimuthal motion, occurring at an instant \edt{when} $U_{\mathrm{eff}}$ becomes small. The simultaneous amplification of both reduced frequencies marks a regime in which the blade motion and the flow response are strongly coupled, and \edt{it} coincides with the onset of \edt{a} rapid \edt{increase} in $\lambda$ \edt{for the turbine to fully self starts}. \edt{These processes do not occur for turbines incapable of self-starting}, although $\alpha_{\mathrm{eff}}$ may undergo large excursions, \fm{its} associated rates of change and kinematic amplification \edt{do not simultaneously undergo their respective} peaks in $k_{\alpha}$ and $k_{\theta}$ that \edt{appears to be} required \edt{for causing a large acceleration of a turbine to take it to a steady $\lambda$}.

\edt{To further explain the unsteady fluid dynamics,} experimental \edt{particle-image velocimetry (PIV) based research for single-bladed turbines, undergoing prescribed rotations \citep{ferreira2007simulating,le2022dynamic}, showed} that the DSV \edt{around the blade followed} a relatively repeatable sequence. Once $\alpha_{\mathrm{eff}}$ exceeds $\alpha_{\mathrm{ss}}$, flow reversal initiates within the boundary layer\edt{,} and the separated leading-edge shear layer rolls up to form a DSV at moderate azimuthal positions, typically around $\theta \approx 30^{\circ}$ to $40^{\circ}$. Continued shear-layer feeding allows the vortex to grow in size and strength while convecting along the suction side of the blade, producing a sharp rise in unsteady aerodynamic loading. As the event progresses, an \edt{oppositely} signed vorticity layer develops between the \edt{blade's} surface and the {vortex} \edt{that interrupts} the feeding process and \edt{cause} \edt{the vortex to detach from the blade} near $\theta \approx 140^{\circ}$ to $160^{\circ}$. Following \edt{this} detachment, the associated loss of suction produces a sharp drop in the pitching moment coefficient $C_m$ \fm{\cite{le2022dynamic,leishman2006principles}}. Due to \edt{the prescribed kinematics of the blade}, these processes are repeatable from cycle to cycle and are commonly reported using phase-averaged flow fields. In the present simulation \edt{framework}, the turbine rotates freely in a continuously accelerating regime with a time-varying  $\lambda$. \edt{Here, all the} blades are \edt{modeled to constitute} a single \edt{rotating turbine. This computational process does not allow the computation of aerodynamic forces and moments for an individual blade. Therefore}, a conventional moment-based dynamic stall analysis for a \edt{single blade is not possible here. It means that we need to rely on the kinematic analysis for an individual blade during the self-starting process to examine its }unsteady aerodynamic behavior associated with dynamic stall. \edt{This explanation provides the justification for our} analysis more focused on kinematics of an individual blade, using \edt{$\alpha_{eff}$ in our earlier discussion}. Dynamic stall behavior is examined only during intervals in which the level of unsteadiness is sufficiently high to support the formation and evolution of coherent dynamic-stall \edt{flow} structures. {As discussed previously}, a reduced frequency of $k=0.05$ {is used as a reference level} beyond which high levels of unsteadiness persist in the flow and unsteady aerodynamic effects become significant. {This value is a commonly accepted indicator of strong unsteady behavior relevant to dynamic stall}. \edt{From the two parameters, $k_{\alpha}$ and $k_{\theta}$, to characterize motion of a VAWT, $k_{\alpha}$ is the primary parameter relevant to dynamic stall, as it directly reflects the rate of change of the effective angle of attack and governs the onset of flow reversal, shear-layer roll-up, and dynamic stall vortex formation \citep{le2021dynamics, le2022dynamic, le2023time}. On the other hand, $k_{\theta}$ characterizes unsteadiness associated with the rate of change of the azimuthal position. Due to this primary difference between the two reduced frequencies, the identification and analysis of dynamic stall events in the present work are based on intervals where $k_{\alpha} > 0.05$. Nevertheless, $k_{\theta}$ is reported to identify phases during which a blade experiences increase in angular speed and to provide information about the overall unsteady state of the turbine.}

\begin{figure}[t] 
    \centering \includegraphics[width=1\linewidth]{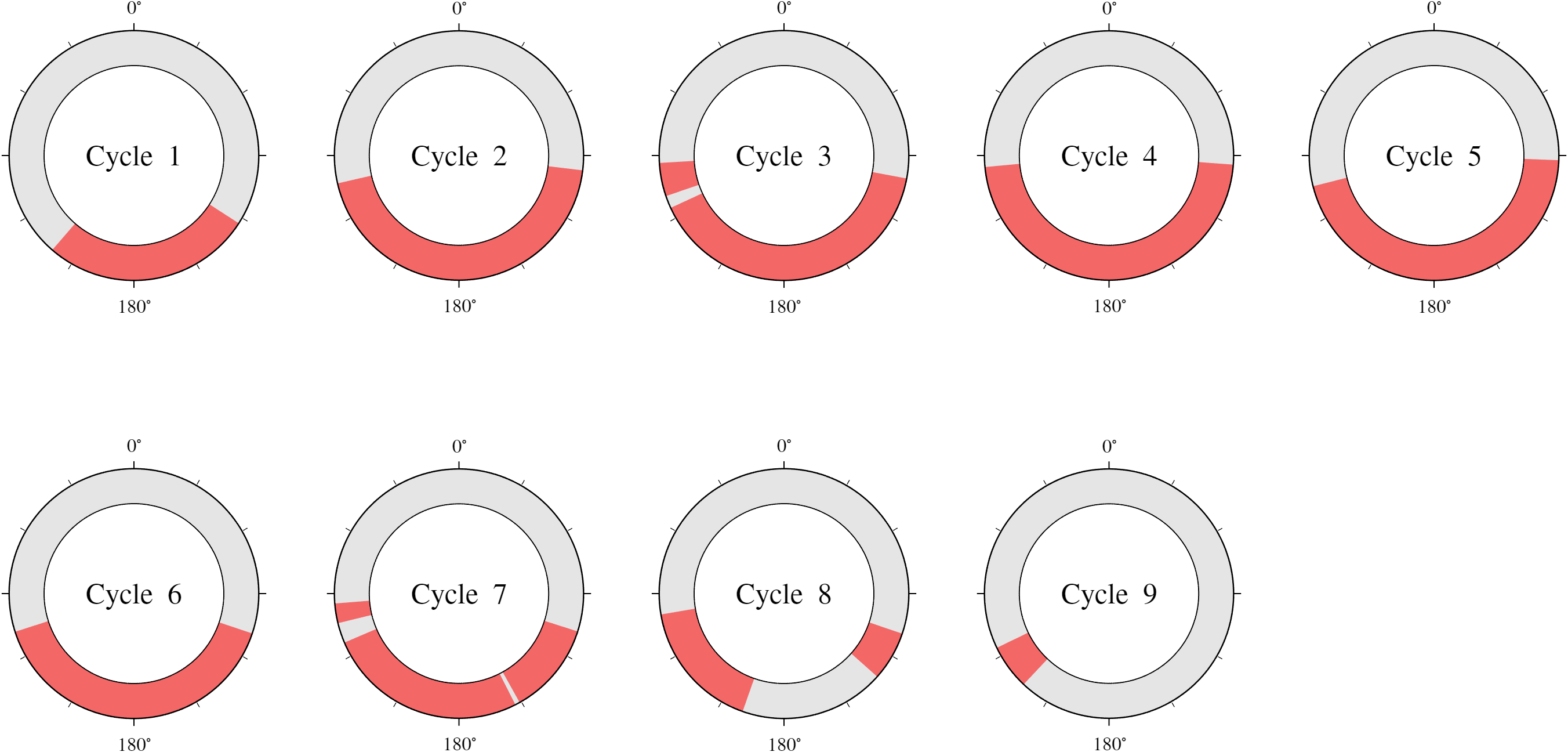} 
    \caption{Annular maps of the pitching reduced frequency for $N=3$, $c=83$~mm, $U_{\infty}=6$~m/s, $k_{\alpha}$, over one revolution in azimuthal angle $\theta$ (0--360$^\circ$) for Cycles~1--9. {Shaded arc segments} indicate azimuthal intervals where $k_{\alpha} > 0.05$, while unshaded portions correspond to $k_{\alpha} \le 0.05$.} 
    \label{fig:kalpha_cycles} 
\end{figure}

\begin{figure}[h!] 
    \centering \includegraphics[width=1\linewidth]{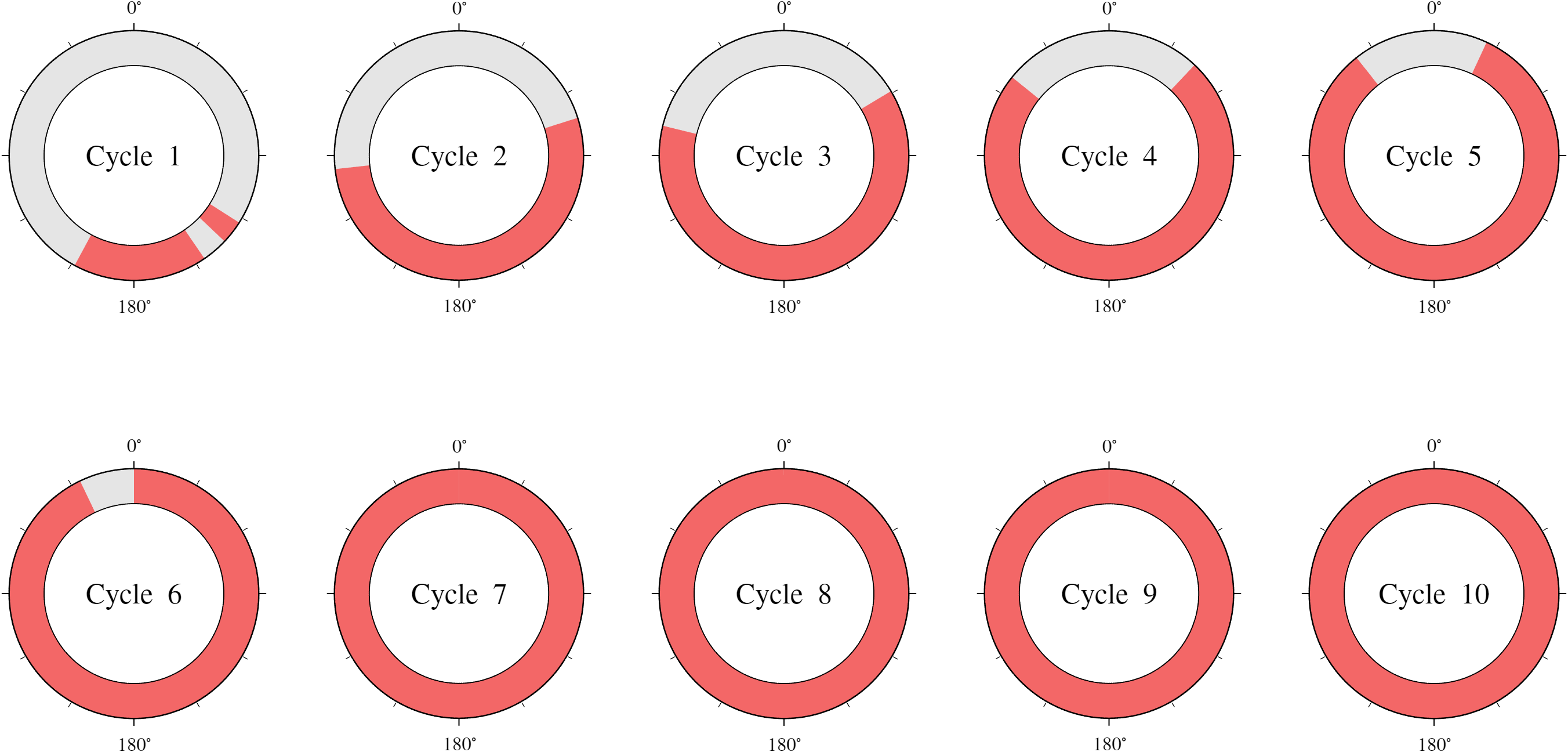} 
    \caption{Annular maps of the azimuthal reduced frequency, $k_{\theta}$, for $N=3$, $c=83$~mm, $U_{\infty}=6$~m/s over one revolution in azimuthal angle $\theta$ (0--360$^\circ$) for Cycles~1--10. {Shaded arc segments} indicate azimuthal intervals where $k_{\theta} > 0.05$.} 
    \label{fig:ktheta_cycles} 
\end{figure}

\edt{Based on a single blade's kinematics, our emphasis now is on determining whether it experiences elevated levels of unsteadiness uniformly throughout its rotational cycle. The plots in Fig.~\ref{fig:kalpha_cycles} demonstrates that it only undergoes these phases at distinct and localized intervals during it rotation.} After Cycle~$9$ \edt{that corresponds} to the onset of self-starting, $k_{\alpha}$ no longer exceeds the reference level. \edt{It directly hints} that dynamic stall\edt{-related} effects diminish significantly once the turbine escapes the dead-band regime and accelerates to higher values of $\lambda$. \edt{Quite oppositely}, $k_{\theta}$ exhibits \edt{an} intermittent behavior during the early cycles but remains continuously elevated once self-starting is achieved, as shown in Fig.~\ref{fig:ktheta_cycles}. \edt{These high values of $k_{\theta}$ demonstrates} sustained azimuthal unsteadiness associated with the increasing rotational speed in the \edt{shaded regimes,} as shown in {Fig.}~\ref{fig:ktheta_cycles}.

\begin{figure}[h!]
  \centering
  \includegraphics[width=1\linewidth]{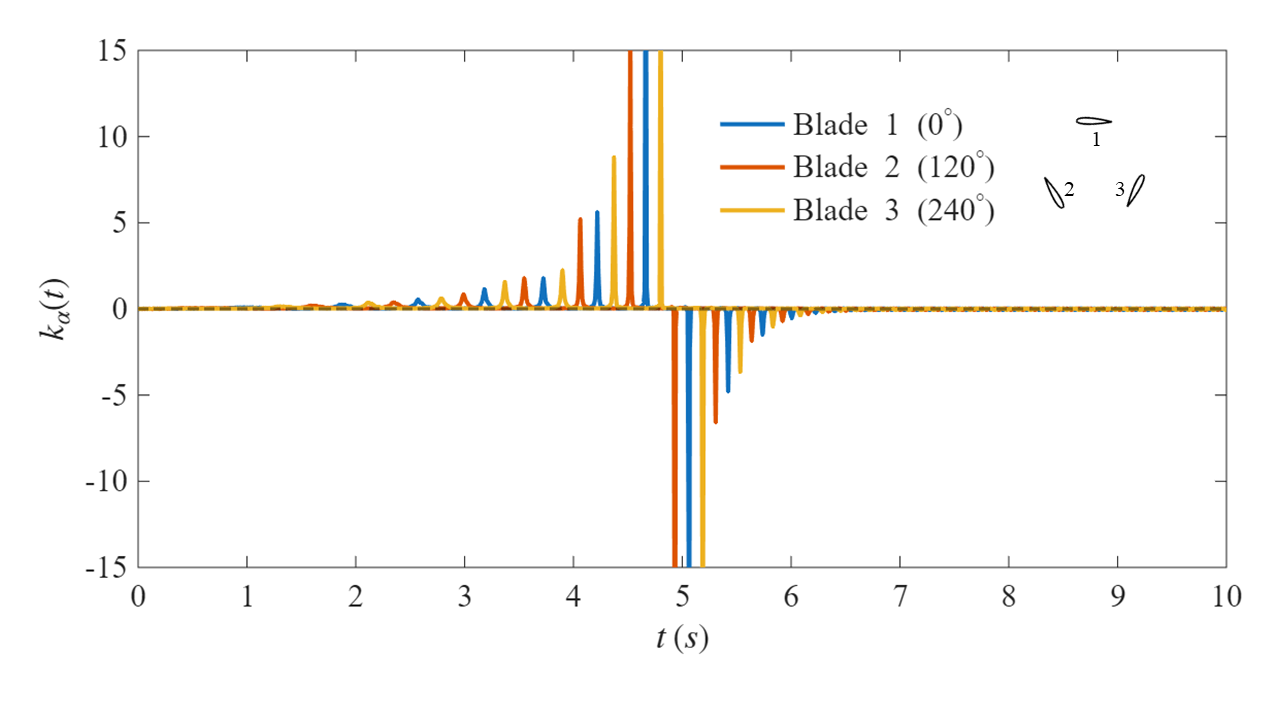}
  \caption{Collective time history of the pitching reduced frequency, $k_{\alpha}(t)$, for $N=3$, $c=83$~mm, $U_{\infty}=6$~m/s}
  \label{fig:kalpha_collective_blades}
\end{figure}

The unsteady behavior of the remaining blades can be \edt{determined} by \edt{incorporating} the appropriate azimuthal phase shift \edt{in the analysis}. As $\lambda$ increases \edt{in successive cycles}, the phase-shifted blades experience larger effective inflow velocities, leading to amplified reduced \edt{frequencies}. This behavior is illustrated in Fig.~\ref{fig:kalpha_collective_blades}, where Blade~3, which is \edt{the} last one in the phase sequence, exhibits very large transient values of $k_{\alpha}$. These large magnitudes arise from the rapid increase in effective inflow velocity with \edt{an} increasing $\lambda$, which amplifies the normalized response even for modest variations in the \fm{$\dot{\alpha}_{eff}$} {pitching rate}.

\begin{figure}[h!]
  \centering
  \includegraphics[width=0.8\linewidth]{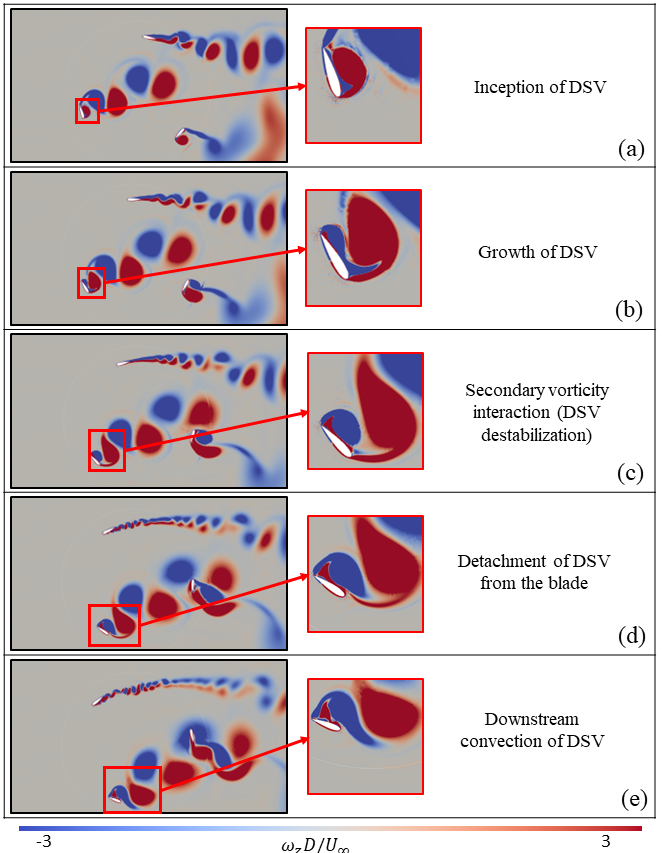}
  \caption{Evolution of the dynamic stall vortex (DSV) during Cycle~4 for $N=3$ at $U_\infty=6~\mathrm{m/s}$ and $c=83~\mathrm{mm}$, visualized using normalized spanwise vorticity, $\omega_z D/U_\infty$}
  \label{fig:dsv_cycle4_sequence_3b_6ms_c83}
\end{figure}

The identified $k_{\alpha}$ intervals therefore represent the most physically meaningful windows for analyzing dynamic stall in \edt{a turbine, experiencing flow-induced rotational acceleration}. Within these intervals, \edt{a blade's} kinematics closely resemble those of an airfoil undergoing dynamic stall under prescribed unsteady motion, as widely reported in the literature \fm{\cite{mccroskey1981phenomenon,ferreira2007simulating,ferreiradynamic, le2022dynamic}}. Importantly, dynamic stall does not occur continuously throughout the startup process, but only during specific phases of the rotation. Similar behaviour \edt{were} reported by \fm{Selvarajoo and Kassim} \cite{selvarajoo2024effects}, who observed that dynamic stall events \edt{occurred} only within limited intervals associated with elevated {pitching} rates. Vorticity analysis performed during a representative $k_{\alpha}$ interval \edt{further} confirms this interpretation. As shown in Fig.~\ref{fig:dsv_cycle4_sequence_3b_6ms_c83}, the classical \edt{sequence of subprocesses involved in} dynamic stall is observed: \edt{(i)} the formation of DSV \edt{in} Fig.~\ref{fig:dsv_cycle4_sequence_3b_6ms_c83}a, (ii) its progressive growth as it is fed by the separated shear layer \edt{in} Fig.~\ref{fig:dsv_cycle4_sequence_3b_6ms_c83}b, (iii) destabilization through the development of oppositely signed vorticity between the vortex and the blade surface \edt{shown in} Fig.~\ref{fig:dsv_cycle4_sequence_3b_6ms_c83}c, and \edt{(iv) its} eventual detachment from the blade \edt{illustrated in} Fig.~\ref{fig:dsv_cycle4_sequence_3b_6ms_c83}d. The detached vortex subsequently convects downstream\edt{, as depicted by} Fig.~\ref{fig:dsv_cycle4_sequence_3b_6ms_c83}e. These observations demonstrate that, despite the freely rotating nature of the turbine, the underlying dynamic stall mechanism remains consistent with that reported for airfoils undergoing prescribed \edt{oscillations}.

\subsection{Comparison of BVI and viscous moment effects in self-starting cases}

\begin{figure}[h!]
    \centering
    \includegraphics[width=0.95\linewidth]{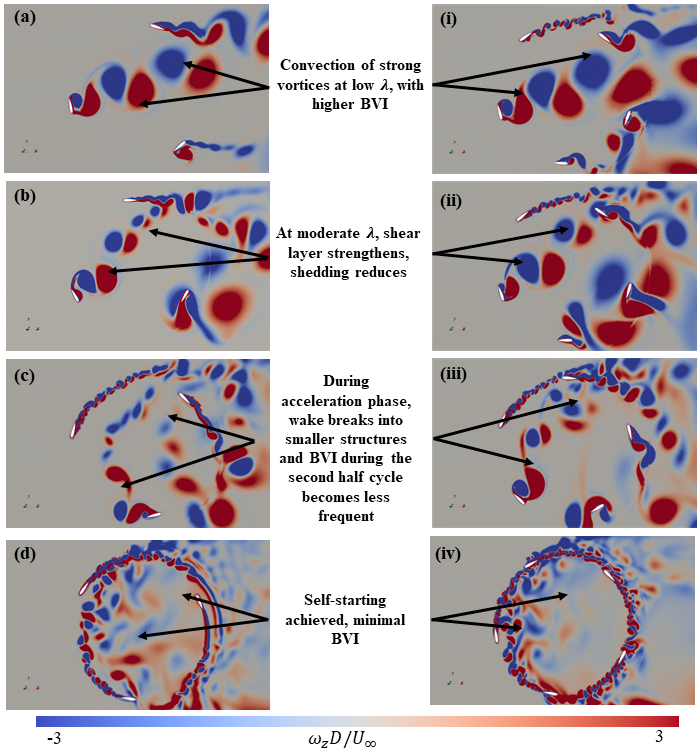}
    \caption{Evolution of vortex development and BVI during self-starting at $c=83~\mathrm{mm}$ and $U_{\infty}=6~\mathrm{m/s}$, comparing $N=3$ (left column, a--d) and $N=5$ (right column, i--iv).}
    \label{fig:bvi}
\end{figure}

To quantify the role of BVI during self-starting, two representative cases \edt{of $3$- and $5$-bladed VAWTs} \edt{with} $c=83~\mathrm{mm}$ are examined at $U_{\infty}=6~\mathrm{m/s}$. The instantaneous vorticity fields are compared in Fig.~\ref{fig:bvi}, where the left column \edt{(Figs.~\ref{fig:bvi}a--\ref{fig:bvi}d)} corresponds to the 3 blade case\edt{,} and the right column \edt{(Figs.~\ref{fig:bvi}i--\ref{fig:bvi}iv)} corresponds to the \edt{$5$-bladed turbine. Please note that both turbine could self start, as identified in Fig.~\ref{fig:summ}}. Both configurations exhibit strong vortex shedding during the early stages, consistent with the initially stalled state, large excursions of $\alpha_{\mathrm{eff}}$, and the absence of an established wake (\fm{Figs.~\ref{fig:bvi}(a)-~\ref{fig:bvi}(i)}). The shed vortices convect to the downstream half of the rotor and interact with \edt{the} blades later in the cycle. This interaction is undesirable\edt{,} because it perturbs the local flow\edt{, approaching the blade}. \edt{It} can reduce the effectiveness of torque production during startup and increase structural loading \edt{ \citep{simao2009visualization,de2021controlling,dunne2015dynamic,ouro2018effect,le2024optimal}}.

\edt{Figures~\ref{fig:bvi}a--\ref{fig:bvi}i} illustrates the early low-$\lambda$ regime, where large, coherent vortices are shed and convected across the rotor. In this phase, the downstream half-cycle contains energetic \edt{flow} structures, increasing the likelihood of strong BVI \edt{there}. At \edt{a} moderate $\lambda$ shown in Fig.~\ref{fig:bvi}b and \ref{fig:bvi}ii, the near-blade shear layer becomes more organized and the most energetic large-scale shedding events become less frequent. During the subsequent acceleration phase \edt{(Figs.~\ref{fig:bvi}c and \ref{fig:bvi}iii)}, the wake breaks into smaller-scale structures and downstream interactions in the second half-cycle become less frequent\edt{. It may happen because} the convected \edt{flow structures} reaching the downwind passage \edt{are} weaker and more fragmented. Once self-starting is achieved (Fig.~\ref{fig:bvi}d and \ref{fig:bvi}iv), the flow field is dominated by a more established wake with reduced large-scale shedding, and only limited downstream interaction is apparent. A clear \edt{distinction} between the two configurations in Fig.~\ref{fig:bvi} is the degree of downstream interaction across the same phases. The \edt{$5$-bladed rotor} occupies a larger fraction of the azimuth at any instant, and therefore has a higher probability of encountering vortical structures convected from the upstream half-cycle. \edt{It} increases the frequency and severity of downstream interactions compared \edt{to those for a $3$-bladed rotor. Consequently,} although additional blades can be beneficial in the start-up stage, the increased \edt{likelihood} of downstream \edt{blade-vortex} interactions can contribute to a lower attainable steady $\lambda$ \edt{when} the turbine transitions out of the dead band.

\begin{figure}[h!]
    \centering
    \includegraphics[width=0.95\linewidth]{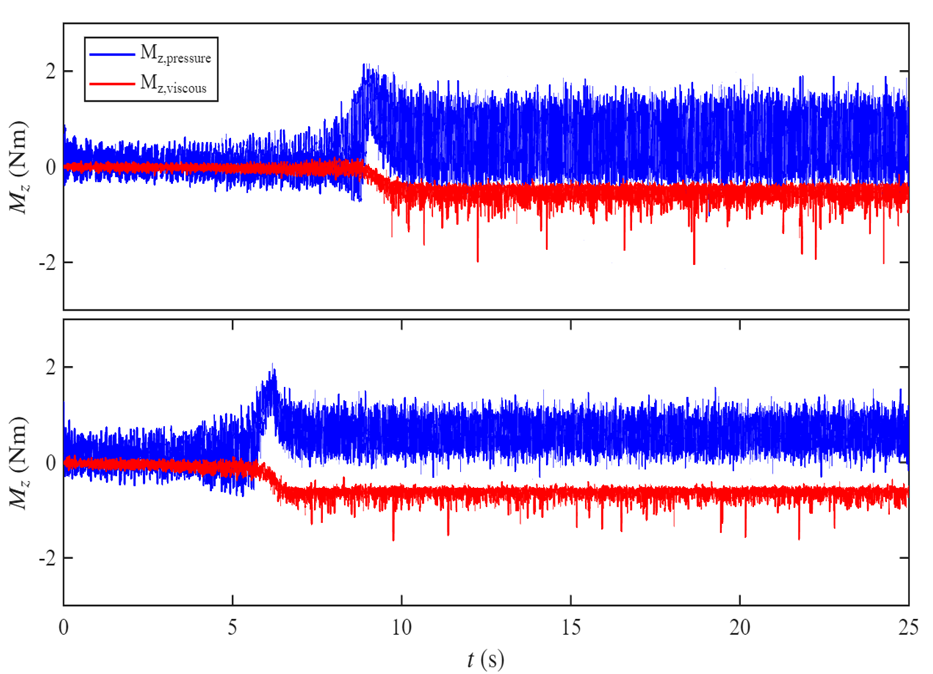}
    \caption{Decomposition of the axial moment into pressure-driven and viscous contributions for \edt{a {$3$-bladed} turbine (upper) and a {$5$-bladed} turbine (lower) with blades of} $c=43~\mathrm{mm}$ at $U_{\infty}=8~\mathrm{m/s}$}
    \label{fig:mom_decomp}
\end{figure}

\edt{Besides,} the same mechanism also explains \edt{the influence of chord on the underlying unsteady fluid dynamics around the rotors} in a general sense. Increasing $c$ \edt{enhances} the spatial footprint of \fm{blade-associated} {separated}\fm{-flow} {regions} and the size of \edt{the} shed vortical structures. \edt{It} also increases the probability of intersecting convected vorticity in the downstream half-cycle \edt{due to a large part of the periphery covered by the blades}. Consequently, even for cases that do self-start, larger chord lengths tend to experience a stronger \edt{BVIs} during recovery and post-startup, which can penalize the final steady $\lambda$. \edt{This interpretation} motivates examining the relative roles of pressure-driven and viscous contributions to the net moment \edt{of the whole turbines around their axis of rotation}. The \edt{two components of the aerodynamic} moment \edt{are} shown in Fig.~\ref{fig:mom_decomp} for \edt{a $3$- and a $5$-bladed turbine with blades of $c=43~\mathrm{mm}$} at $U_{\infty}=8~\mathrm{m/s}$. These two cases are selected because, for each blade count, they reach the maximum $\lambda_{\mathrm{steady}}$ following the transient acceleration phase, and therefore provide a consistent basis for comparing the torque balance at the highest attainable operating state. From Fig.~\ref{fig:mom_decomp}, the mean viscous contribution to $M_z$ is more negative for the \edt{$5$-bladed} turbine than for \edt{its $3$-blades counterpart. Their time-averaged values are $-0.48$ and $-0.33$}, respectively. As a result, a larger fraction of the available driving moment \edt{primarily relates to} skin-friction and near-wall shear \edt{effects}, \edt{which reduces} the net torque available to sustain rotation \edt{for the turbine}. Consequently, the \edt{$5$-bladed VAWT} attains a lower $\lambda_{\mathrm{steady}}$ than the \edt{$3$-bladed one} under the same operating \edt{flow} parameters. During the early startup and dead-band evolution, the viscous contribution remains small relative to the pressure-driven moment, while the pressure-driven moment exhibits strong fluctuations and frequent sign changes consistent with repeated separation/recovery and interaction events. As the \edt{turbines} transitions into sustained acceleration, the pressure-driven moment becomes persistently positive and increases sharply. After sustained rotation is established, the viscous contribution becomes clearly non-negligible and remains predominantly negative, \edt{which indicates} that viscous stresses act as an \text{aerodynamic braking mechanism} \edt{in order to} limit further \edt{increase in the angular speed of the VAWT}.

\section{Conclusions}
\label{sec:conclusions}

This study examines the influence of $c$ and $N$ on the self-starting behavior and steady operating state of a Darrieus-type VAWT using two-dimensional URANS simulations of freely accelerating rotors. EC and ES configuration families are analyzed to isolate the individual and combined effects of $c$ and $N$ on startup dynamics, unsteady aerodynamics, and the attainable $\lambda_{\mathrm{steady}}$.

The results indicate that geometric parameters influence self-starting indirectly, while the decisive factor is the persistence of organized unsteady aerodynamic loading capable of producing cumulative positive torque. For both the 3 blade and 5 blade configurations at constant $R$ and $J$, self-starting occurs only above $c_{\mathrm{crit}}$, below which the turbine remains trapped in the dead-band regime with $\lambda_{\mathrm{steady}} < 1$. The existence of $c_{\mathrm{crit}}$ therefore defines a clear geometric threshold separating non-self-starting and self-starting behavior. Increasing the blade number shifts this threshold to larger values of $c$, indicating that a greater chord is required to sustain the level of unsteadiness necessary for startup. In the ES family, this shift becomes sufficiently strong that self-starting is suppressed entirely, demonstrating that increasing $N$ under comparable operating conditions is detrimental to sustained rotation.

The dynamic stall analysis shows that, in a freely accelerating VAWT, dynamic stall is not sustained throughout startup but occurs intermittently over localized azimuthal intervals associated with elevated pitching unsteadiness. The reduced frequency $k_{\alpha}$ emerges as the primary parameter governing dynamic stall activity, as it directly reflects the rate of change of $\alpha_{\mathrm{eff}}$ and controls the onset of flow reversal, shear-layer roll-up, and dynamic stall vortex formation. Exceeding the unsteady limit alone is insufficient to induce dynamic stall, rather, sustained intervals of elevated $k_{\alpha}$ are required to allow the separated shear layer to evolve into a coherent vortex structure. During the early startup and dead-band phases, intermittent dynamic stall events are observed, while their magnitude and duration diminish as the turbine accelerates and transitions toward sustained rotation. Once self-starting is achieved, $k_{\alpha}$ remains below the unsteady threshold, indicating that dynamic stall is suppressed at higher operating states, even though $k_{\theta}$ may remain elevated. These results demonstrate that dynamic stall in a freely rotating VAWT is governed by the temporal organization of pitching-rate unsteadiness rather than by instantaneous angle of attack alone, and that reduced-frequency measures provide a physically meaningful framework for identifying and analyzing dynamic stall activity during startup.

Vorticity contours analysis demonstrate that vortices generated during the upwind half-cycle convect into the downwind half and interact with subsequent blades, producing resistive loading through BVI. The frequency and severity of this interaction increase with blade number and chord length, providing a consistent physical explanation for the reduction in $\lambda_{\mathrm{steady}}$ observed for 5 blade turbines and for larger-chord configurations that do self-start.

The moment decomposition shows that viscous effects play an active role in the self-starting process. While viscous contributions are small during the earliest stages, they become dynamically significant during the acceleration phase and remain predominantly negative once sustained rotation is established. At the highest attainable operating states, the 5 blade turbine exhibits a more negative mean viscous contribution than the 3 blade turbine, indicating stronger viscous dissipation and reduced net torque availability. This increased viscous dominance directly limits the attainable $\lambda_{\mathrm{steady}}$.

From a design perspective, the results demonstrate a clear startup--performance trade-off. Increasing $c$ or $N$ promotes early-stage unsteady loading and facilitates escape from the dead band, but simultaneously increases BVI and viscous losses that limit the steady operating state. The identification of a critical chord length and the use of reduced-frequency metrics provide practical tools for diagnosing startup capability and dynamic stall activity. These tools offer a physically grounded framework through which future geometric modifications, such as changes in chord distribution, blade count, or airfoil selection, can be assessed and guided toward improved self-starting performance without compromising steady-state efficiency.

Although the present results are obtained from two-dimensional URANS simulations, the observed relationships between critical chord length, sustained unsteadiness, reduced-frequency behaviour, BVI, and viscous moment contribution remain consistent across the parameter space considered. The findings therefore provide design relevant insight into the mechanisms governing self-starting and steady operation in lift-driven VAWTs, and establish a foundation for future three-dimensional and experimentally coupled investigations.

\section*{Acknowledgment}
MSU Khalid acknowledges the funding support from Lakehead University through the startup grant and the Natural Sciences and Engineering Research Council of Canada (NSERC) through the Discovery grant program. \edt{F. Muhammad is thankful to Lakehead University for the graduate scholarship and for the support through the Ontario Graduate Scholarship.} The simulations reported in this work were performed on the supercomputing clusters administered and managed by the Digital Research Alliance of Canada. 


\vskip6pt

\bibliography{References_Faisal}

    \end{document}